\documentclass{ifacconf}

\usepackage{graphicx}      

\usepackage{natbib}        
\usepackage{multicol}
\usepackage{amsmath}
\usepackage{amssymb}
\usepackage{accents}
\usepackage{algorithm}
\usepackage{etoolbox}
\usepackage{xcolor}
\usepackage{color,soul}
\let\classAND\AND
\let\AND\relax
\usepackage{algorithmic}

\let\AND\classAND

\AtBeginEnvironment{algorithmic}{\let\AND\algoAND}
\begin{document}
\begin{frontmatter}
\title{Joint Stabilization and Regret Minimization through Switching in Over-Actuated Systems (extended version)\thanksref{footnoteinfo}} 

\thanks[footnoteinfo]{              }

\author[First]{Jafar Abbaszadeh Chekan} 
\author[Second]{Kamyar Azizzadenesheli} 
\author[First]{C\'edric Langbort} 

\address[First]{Coordinated Science Lab, University of Illinois at Urbana-Champaign, IL 61820 USA (e-mail: langbort \& jafar2@illinois.edu).}
\address[Second]{Department of Computer Science, Purdue University, IN 47907 USA (e-mail: kamyar@purdue.edu)}

\begin{abstract}                
Adaptively controlling and minimizing regret in unknown dynamical systems while controlling the growth of the system state is crucial in real-world applications. In this work, we study the problem of stabilization and regret minimization of linear over-actuated dynamical systems. We propose an optimism-based algorithm that leverages possibility of switching between actuating modes in order to alleviate state explosion during initial time steps. We theoretically study the rate at which our algorithm learns a stabilizing controller and prove that it achieves a regret upper bound of $\mathcal{O}(\sqrt{T})$.
\end{abstract}


\end{frontmatter}

\section{Introduction}

The past few years have witnessed a growing interest in an online learning-based Linear Quadratic (LQ) control,
in which an unknown LTI system is controlled while guaranteeing a suitable scaling of the regret (defined as the average difference between the closed-loop quadratic cost and the best achievable one, with the benefit of knowing the plant's parameters)  over a desired horizon $[0,T]$.

Table 1 summarizes the regret's scaling achieved by several recent works in the literature (\cite{abbasi2011regret, mania2019certainty, cohen2019learning, lale2020explore}). As can be seen, the bounds scale like $\sqrt{T}$, but also include an exponential term in $m$ (the dimension of the plant's state \textit{plus inputs}) when an initial stabilizing controller is unavailable. Recently, \cite{chen2020black} have shown that this dependency is unavoidable in this setting, at least for an initial exploration time period, by providing a matching lower bound. On closer inspection, this undesirable dependency stems from an exponential growth of the system's state during the initial period of learning, which eventually contributes a $\mathcal{O}(m^m)$ term like that appearing on the last row of Table 1. Apart from negatively impacting regret, this transient state growth can also be damageable if, e.g., the linear plant to be controlled is in fact the linearization of a nonlinear process around some particular equilibrium, and the state can be driven outside the neighborhood of this equilibrium where the linearization is adequate.


A natural idea to try and partially alleviate this effect in over-actuated systems is to reduce the ambient dimension $``m"$ during the initial period by employing only a subset of all available actuators, before potentially switching to a different mode to simultaneously learn and control the plant. The goal of this paper is to show that this approach yields the desired result (a lower bound on the plant's state and the regret than in the presence of all actuators) and to provide a rigorous proof of the achieved $\mathcal{O}((n+d_i)^{(n+d_i)})$ bound (where $d_i$ is number of actuators during initial exploration). We note that, while the idea is conceptually simple, obtaining these rigorous guarantees is not completely straightforward, and requires a revisit of the tools of \cite{lale2020explore} to overcome a potential challenge: ensuring that, at the end of the period when a strict subset of actuators is used, all entries of the B-matrix (including those corresponding to unused actuators) are sufficiently well-learned for the closed-loop to become stable and the regret to remain appropriately bounded. This can only be achieved by adding additional exploratory noise with respect to the approach of \cite{lale2020explore} and results in an additional linear term in the regret bound.

\begin{table}[h!]
\scriptsize
\caption{Overview of Prior Works}
\centering
\begin{tabular}{c c c}

\hline \hline
Algorithm  & Regret& *\\ [1ex]
\hline
\cite{abbasi2011regret}&  $\mathcal{O}(m^m\sqrt{T})$ & No \\
\cite{mania2019certainty} & $\mathcal{O}(poly(m)\sqrt{T})$  & Yes \\
\cite{cohen2019learning} &  $\mathcal{O}(poly(m)\sqrt{T})$  & Yes  \\
\cite{lale2020explore}&  $\mathcal{O}(m^m)+\mathcal{O}(poly(m)\sqrt{T})$  & No \\
\hline
\end{tabular}
\end{table}
* requirement for initial stabilizing controller.

While this idea of leveraging over-actuation can generally be applied to any model-based type of online learning algorithms presenting the exponential scaling mentioned above, we focus on the special kind of methods based on ``Optimism in the Face of Uncertainty (OFU)''.


OFU type of algorithms, which couple estimation and control design procedures, have shown their ability to outperform the naive certainty equivalence algorithm. \cite{campi1998adaptive} propose an OFU based approach to address the optimal control problem for LQ setting with guaranteed asymptotic optimality. However, this algorithm only guarantees the convergence of average cost to that of the optimal control in the limit but does not provide any bound on the measure of performance loss in finite time. \cite{abbasi2011regret} propose a learning-based algorithm to address
the adaptive control design of LQ control system in finite time with worse case regret bound of $\mathcal{O}(\sqrt{T})$ with exponential dependence in dimension. Using $l_2$-regularized least square estimation, they construct a high-probability confidence set around unknown parameters of the system and designed an algorithm that optimistically plays with respect to this set. Along this line, many works attempt to get rid of the exponential dependence with further assumption, e.g. highly sparse dynamics (see \cite{ibrahimi2012efficient}) or access to a stabilizing controller (see \cite{cohen2019learning}). Furthermore, \cite{faradonbeh2017optimism} propose an OFU-based learning algorithm with mild assumptions and $\mathcal{O}(\sqrt{T})$ regret. This class of algorithms was extended by  \cite{lale2020regret,lale2020adaptive} to LQG setting where there is only partial and noisy observations on the state of system. In addition, \cite{lale2020explore} propose an algorithm with more exploration for both controllable and stabilizable systems. 

The remainder of the paper is organized as follows:
Section \ref{sec:prelim} reviews the preliminaries, assumptions and presents the problem statement and formulation. Section
\ref{problem statement} overviews the proposed initial exploration (IExp), stabilizing OFU (SOFUA) algorithms and discusses in detail how to choose best actuating mode for the initial exploration purpose.  Furthermore, in Section \ref{problem statement} the performance analysis (state norm upper-bound and regret bound) is given while leaving the details of the analysis to Section \ref{sec:analysis} and technical theorems and proofs of analysis to Section
\ref{appendix}. Numerical experiments are given in Section \ref{simulation}. Finally, Section \ref{conclusion} summarizes the paper's key contributions.

\section{Assumptions and Problem Formulation}\label{sec:prelim}

Consider the following linear time invariant dynamics and the associated cost functional given by:
\begin{subequations}
	\begin{align}
		x_{t+1} &=A_{*}x_{t} + B_{*}u_{t}+\omega_{t+1}\label{eq:dyn}\\
		c_{t} &=x_{t}^\top Q_*x_{t} + u_{t}^\top R_*u_{t}\label{eq:obs}
	\end{align}
\end{subequations}
where the plant and input matrices $A_{*}\in  \mathbb{R}^{n \times n}$ and  $B_{*}\in  \mathbb{R}^{n \times d}$ are initially unknown and have to be learned and $(A_*,B_*)$ is controllable. $Q_*\in  \mathbb{R}^{n \times n} $ and $R_*\in  \mathbb{R}^{d \times d} $ represent known and positive definite matrices. $\omega_{t+1}$ denotes process noise, satisfying the following assumption. 

\begin{assum}

\label{Assumption 1}
There exists a filtration $\mathcal{F}_{t}$ such that

$(1.1)$ $\mathbb{E}[\omega_{t+1}\omega_{t+1}^\top|\;\mathcal{F}_{t}]=\bar{\sigma}_{\omega}^2I_{n}$ for some $\bar{\sigma}_{\omega}^2>0$;

$(1.2)$ $\omega_{t}$ are component-wise sub-Gaussian, i.e., there exists $\sigma_{\omega}>0$ such that for any $\gamma \in \mathbb{R}$ and $j=1,2,...,n$
\begin{align*}
\mathbb{E}[e^{\gamma\omega_{j}(t+1)}|\;\mathcal{F}_{t}]\leq e^{\gamma^2\sigma_{\omega}^2/2}.
\end{align*}
\end{assum}



The problem is designing a sequence $\{u_t\}$ of control inputs such that the regret $\mathcal{R}_T$ defined by

\begin{align}
	\mathcal{R}_T =\sum _{t=1}^{T} \bigg( x_{t}^\top Q_*x_{t} + u^\top_{t} R_* u_{t}-J_*(\Theta_*, Q_*, R_*)\bigg)\label{eq:Reg} 
\end{align}

achieves a desired specification which scales sublinearly in $T$. The term $J_*(\Theta_*, Q_*, R_*)$ in (\ref{eq:Reg}) where $\Theta_*=(A_*\; B_*)^\top$ denotes optimal average expected cost. For LQR setting with controllable pair $(A,\;B)$ we have $J_*(\Theta,Q,R)=\bar{\sigma}_{\omega}^2 trace( P(\Theta,Q,R))$, where $P(\Theta,Q,R)$ is the unique solution of discrete algebraic riccati equation (DARE) and the average expected cost minimizing policy has feedback gain of 
\begin{align*}
K(\Theta,Q,R)= -(B^\top P(\Theta, Q,R)B+R)^{-1}B^\top P(\Theta, Q, R)A.
\end{align*}






While the regret's exponential dependency on system dimension appears in the long-run  in \cite{abbasi2011regret} the recent results of \cite{mania2019certainty} on the existence of a stabilizing neighborhood, makes it possible to design an algorithm that only exhibits  this dependency during an initial exploration phase (see \cite{lale2020explore}).  

After this period, the controller designed for any estimated value of the parameters is guaranteed to be stabilizing and the exponentially dependent term thus only appears as a constant in overall regret bound. As explained in the introduction, this suggests using only a subset of actuators during initial exploration to even further reduce the guaranteed upperbound on the state.

In the remainder of the paper, we pick the best actuating mode (i.e. subset of actuators) so as to minimize the state norm upper-bound achieved during initial exploration and characterize the needed duration of this phase for all system parameter estimates to reside in the stabilizing neighborhood. This is necessary to guarantee both closed loop stability and acceptable regret, and makes it possible to switch to the full actuation mode. 

Let $\mathbb{B}$ be the set of all columns, $b^i_*$ ($i\in \{1,...,d\}$) of $B_*$. An element of its power set $2^{\mathbb{B}}$ is a subset $\mathcal{B}_{j}$ $j\in \{1,..., 2^d\}$ of columns corresponding to a submatrix $B_*^j$ of $B_*$ and mode $j$. For simplicity, we assume that $B^1_*=B_*$ i.e., that the first mode contains all actuators. Given this definition we write down different actuating modes dynamics with extra exploratory noise as follows

\begin{align}
	x _{t+1} ={\Theta^i_{*}}^\top z_{t}+B_*\nu_t+\omega_{t+1}, \quad z_t=\begin{pmatrix} x_t \\ u^i_t \end{pmatrix}. \label{eq:dynam_by_theta} 
\end{align}
where $\Theta^i_*=(A_*,B^i_*)^\top$ is controllable. 

The associated cost with this mode is
\begin{align}
		c^i_{t} &=x_{t}^\top Q_* x_{t} + {u^i_t}^\top R^i_* u^i_t \label{eq:obsswitch}
\end{align}
where $R_*^i\in \mathbb{R}^{d_i\times d_i}$ is a block of $R_*$ which penalizes the control inputs of the actuators of mode $i$. 


We have the following assumption on the modes which assists us in designing proposed strategy.





\begin{assum}(Side Information) \label{Assumption2}
\begin{enumerate}
\item  There exists $s^i$ and $\Upsilon_i$ such that $\Theta_{*}^i \in \mathcal{S}^i_c$ for all modes $i$ where
\begin{align*}
  \mathcal{S}^i_c =&\{{\Theta^i} \in R^{(n+d_i)\times n}  \mid trace({\Theta^i}^\top \Theta^i)\leq ({s^i})^2,\\
 &\text{$(A,B^i)$ is 
 controllable,}\\
 &  
\|A+B^iK(\Theta^i,Q_*,R^i_*)\|\leq  \Upsilon_i<1  \\
&
\text{and $(A,M)$ is observable,} \text{where $Q=M^\top M$} \}.
\end{align*}
.
\item There are known positive constants  $\eta^i$, $\vartheta_i$, $\gamma^i$ such that $\|B_*^i\|\leq \vartheta_i$,
\begin{align}
&\sup_{\Theta^i\in \mathcal{S}^i}\|A_*+B^i_*K({\Theta}^i,Q_*,R^i_*)\|\leq \eta^i \label{Assum3_1}
\end{align}
and
\begin{align}
J_*(\Theta_*^{i},Q_*,R^i_*)-J_*(\Theta_*,Q_*,R_*)\leq \gamma^i. \label{Assum3_4}
\end{align}	
for every mode $i$.
\end{enumerate}
\end{assum}

By slightly abusing notation, we drop the superscript label for the actuating mode 1 (e.g. $\Upsilon_1=\Upsilon$, $s^1=s$, and $\mathcal{S}_c^1=\mathcal{S}_c$). It is obvious that $s^i\leq s$ $\forall i$. 

Note that the item (1) in Assumption \ref{Assumption2} is typical in the literature of OFU-based algorithms (see e.g., \cite{abbasi2011regret,lale2020explore}) while (2) in fact always holds in the sense that $\sup_{\Theta^i\in \mathcal{S}^i}\|A_*+B^i_*K({\Theta}^i,Q_*,R^i_*)\|$ and $J_*(\Theta_*^{i},Q_*,R^i_*)-J_*(\Theta_*,Q_*,R_*)$ are always bounded (see e.g., \cite{abbasi2011regret,lale2020explore}). The point of (2), then, is that upper-bounds on their suprema are available which can in turn be used to bound regret explicitly. The knowledge of these bounds does not affect Algorithms 1 and 2 but their value enters Algorithm 3 for determination of the best actuating mode and the corresponding exploration duration. In that sense "best actuating mode" should be understood as "best given the available information".

Boundedness of $S^i_c$'s implies boundedness of $P(\Theta^i,Q_*,R_*^i)$ with a finite constant $D_c^i$ (see \cite{anderson1971linear}), (i.e., $D_c^i=\sup \{\left\lVert P(\Theta^i,Q_*,R_*^i)\right\rVert \mid\Theta^i \in \mathcal{S}^i_c \}$). We define $D=\max_{i\in \mathcal{B}^*} D^i$. Furthermore, there exists $\kappa^i_c>1$ such that $\kappa^i_c=\sup \{\left\lVert K(\Theta^i, Q_*, R_*^i)\right\rVert \mid\Theta ^i\in \mathcal{S}^i_c \}$.

Recalling that the set of actuators of mode $i$ is $\mathcal{B}_{i}$, we denote its complement by $\mathcal{B}_i^c$ (i.e. $\mathcal{B}_i\cup \mathcal{B}_i^c=\{1,...,d\}$). Furthermore, we denote the complement of control matrix $B_*^i$ by $\bar{B}^i_*$.

If some modes fail to satisfy Assumption
\ref{Assumption2} they can simply be removed from the set $2^{\mathbb{B}}$ without affecting algorithm or the derived guarantees.

\section{Overview of Proposed Strategy}\label{problem statement}

In this section, we propose an  algorithm in the spirit of that first proposed by \cite{lale2020explore} which leverages actuator redundancy in the "more exploration" step to avoid blow up in the state norm while minimizing the regret bound. We break down the strategy into two phases of initial exploration, presented by Algorithm (IExp), and optimism (Opt), given by SOFUA algorithm. 

The IExp algorithm, which leverages exploratory noise, is deployed in the actuating mode $i^*$ for duration $T_c^{i^*}$ to reach a stabilizing neighborhood of the full-actuation mode and alleviate state explosion while minimizing regret. 

Afterwards, Algorithm 2 which leverages all the actuators comes into play. This algorithm has the central confidence set, given by the Algorithm 1, as an input. The best actuating mode $i^*$ that guarantees minimum possible state norm upper-bound and initial exploration duration $T^{i^*}_c$ is determined by running Algorithm 3 at the subsection \ref{deterBest}.

\begin{algorithm} 
	\caption{\small Initial Exploration (IExp) \normalsize} \label{alg:IExp}
		\begin{algorithmic}[1]
		\STATE \textbf{Inputs:}$T^{i^*}_c$$\,s^{i^*}>0,$$\,\delta>0,$$\, \sigma_{\omega},\, \sigma_{\nu}\,,\lambda>0$
		\STATE set $V^{i^*}_0=\lambda I$, $\hat{\Theta}^{i^*}=0$ 
		\STATE $\tilde{\Theta}^i_0=\arg\min_{\Theta \in \mathcal{C}^{i^*}_0(\delta)\cap S^i}\,\, J(\Theta^i,Q_*,R_*^i)$
		\FOR {$t = 0,1,..., T^{i^*}_c$}
		\IF {$\det(V^{i^*}_t)> 2 \det(V^{i^*}_{\tau})$ or $t=0$} 
		\STATE Calculate $\hat{\Theta}^{i^*}_t$ by (\ref{eq:LSE_Solf}) and set $\tau=t$

		\STATE Find $\tilde{\Theta}^{i^*}_t$ by (\ref{eq:nonconvexOpt}) for $i=i^*$
		\ELSE
		\STATE $\tilde{\Theta}^{i^*}_t=\tilde{\Theta}^{i^*}_{t-1}$
		\ENDIF 
		\STATE For the parameter $\tilde{\Theta}^{i^*}_t$ solve the Ricatti equation and find $u^{i^*}_t=K(\tilde{\Theta}^{i^*}_t,Q_*, R_*^{i^*})x_t$
		\STATE Construct $\bar{u}^{i^*}_t$ using (\ref{eq:controlAlg}) and apply it on the system $\Theta_*$ (\ref{eq:dynam_by_theta2}) and observe new state $x_{t+1}$.
		\STATE Using $u^{i^*}_t$ and $x_t$ form $z^{i^*}$ and Save $(z^{i^*}_t,x_{t+1})$ into dataset
		\STATE Set $V^{i^*}_{t+1}=V^{i^*}_t+z^{i^*}_t{z^{i^*}_t}^\top$ and form $\mathcal{C}^{i^*}_{t+1}$
		\STATE using $\bar{u}^i_t$ and $x_t$ form $\bar{z}^i_t$
		\STATE Form $(\bar{z}^i_t,x_{t+1})$
		\STATE Set $V_{t+1}=V_t+\bar{z}^i_t {\bar{z}_t^i}^\top$ and form $\mathcal{C}_{t+1}$\ENDFOR
		\STATE Return $V_{T_c+1}$ and corresponding $\mathcal{C}_{T_c}$ 
	\end{algorithmic}
	\end{algorithm}

\begin{algorithm} 
	\caption{Stabilizing OFU Algorithm (SOFUA)} \label{alg:SOFUA}
		\begin{algorithmic}[1]
		\STATE \textbf{Inputs:}$T,$$\,S>0,$$\,\delta>0,$$\,Q\,, L,\, V_{T_c},\, \mathcal{C}_{T_c},\, \hat{\Theta}_{T_c}$
		\STATE $\tilde{\Theta}_{T_c}=argmin_{\Theta \in \mathcal{C}_{T_c}(\delta)\cap S}\,\, J(\Theta)$
		\FOR {$t = T_c,T_c+1,T_c+2,...$}
		\IF {$\det(V_t)> 2 \det(V_{\tau})$ or $t=T_c$} 
		\STATE Calculate $\hat{\Theta}_t$ by (\ref{eq:LSE_Solf}) and set $\tau=t$

		\STATE Find $\tilde{\Theta}_t$ by (\ref{eq:nonconvexOpt}) for $i=1$
		\ELSE
		\STATE $\tilde{\Theta}_t=\tilde{\Theta}_{t-1}$
		\ENDIF 
		\STATE For the parameter $\tilde{\Theta}_t$ solve Ricatti and calculate $\bar{u}_t=K(\tilde{\Theta}_t, Q_*,R_*)x_t$
		\STATE Apply the control on $\Theta_*$ and observe new state $x_{t+1}$.
		\STATE Save $(z_t,x_{t+1}) $ into dataset
		\STATE $V_{t+1}=V_t+z_tz_t^\top$
		\ENDFOR
	\end{algorithmic}
	\end{algorithm}
\subsection{Main steps of Algorithm 1}

\subsubsection{Confidence Set Contruction}

In IExp phase we add an extra exploratory Gaussian noise $\nu$ to the input of all actuators even those not in actuators set of mode $i$. Assuming that the system actuates in an arbitrary mode $i$, the dynamics of system, used for confidence set construction (i.e. system identification), is written as
\begin{align}
	x _{t+1} ={\Theta^i _{*}}^\top \underline{z}^i_{t}+\bar{B}^i_*\bar{\nu}^i_t+\omega_{t+1}, \quad \underline{z}^i_{t}=\begin{pmatrix} x_t \\ \underline{u}^i_t \end{pmatrix}. \label{eq:dynam_by_theta3} 
\end{align}

in which $\bar{B}^i_*\in \mathbb{R}^{d-d_i}$ and $\underline{u}^i_{t}=u^i_t+\nu_t(\mathcal{B}_i)$,  and $\bar{\nu}^i_t=\nu_t(\mathcal{B}_i^c)$ where, if $\nu_t \in \mathbb{R}^{d}$ and $\mathcal{N}\subset\mathbb{B}$, the vector $\nu (\mathcal{N})\in \mathbb{R}^{card(\mathcal{N})}$ is constructed by only keeping the entries of $\nu_t$ corresponding to the index set of elements in $\mathcal{N}$. Note that (\ref{eq:dynam_by_theta3}) is equivalent to (\ref{eq:dynam_by_theta}) but separates used and unused actuators.

By applying self-normalized process, the least square estimation error, $e(\Theta^{i})$ can be obtained as:
	\begin{align}
	\nonumber & e(\Theta^{i})= \lambda  \operatorname{Tr}({\Theta^{i}}^\top\Theta^{i})\\
	&+\sum _{s=0}^{t-1} \operatorname{Tr} \big((x_{s+1}-{\Theta^{i}}^\top \underline{z}^{i}_{s})(x_{s+1}-{\Theta^{i}}^\top \underline{z}^{i}_{s})^\top)\big)  \label{eq:LSE_op}
	\end{align}
with regularization parameter $\lambda$. This yields the $l^{2}$-regularized least square estimate:

\begin{align}
	\hat{\Theta}^i_t &=\operatorname*{argmin_{\Theta^{i}}} e(\Theta^{i})=({\underline{Z}_t^{i}}^\top \underline{Z}_t^{i}+\lambda I)^{-1}{\underline{Z}_t^{i}}^\top X_t
	\label{eq:LSE_Solf}
\end{align}

where $\underline{Z}_t^{i}$ and $X_t$ are matrices whose rows are ${\underline{z}^{i}_{0}}^\top,..., {\underline{z}^{i}_{t-1}}^\top$ and $x_{1}^\top,...,x_{t}^\top$, respectively.
Defining covariance matrix $V^{i^*}_{t}$ as follows:
\begin{align*}
V^{i}_{t}=\lambda I + \sum_{s=0}^{t-1} \underline{z}^{i}_{s}{\underline{z}^{i}_{s}}^\top=\lambda I +{\underline{Z}_t^{i}}^\top \underline{Z}_t^{i},
\end{align*}

it can be shown that with probability at least $(1-\delta)$, where $0<\delta<1$, the true parameters of system $\Theta^i_*$ belongs to the confidence set defined by (see  \ref{thm:Conficence_Set_Attacked}): 
\begin{align}
	\nonumber\mathcal{C}^{i}_{t}(\delta)& =\{{\Theta^i}^\top \in R^{n \times (n+d_i)}  \mid \\  & \nonumber\operatorname{Tr}((\hat{\Theta}^{i}_{t}-\Theta^{i})^\top V_{t}^i(\hat{\Theta}^{i}_{t}-\Theta^{i}))\leq \beta^{i}_{t}(\delta)  \}, \\
	\nonumber\beta^{i}_t(\delta) &=\bigg(\lambda^{1/2}s^i+\sigma_{\omega}\sqrt{2n\log(n\frac{\det(V^{i}_{t})^{1/2}\det(\lambda I)^{-1/2}}{\delta}})\\
	&+\|\bar{B}_*^i\|\sigma_{\nu}\sqrt{2d_i\log(d_i\frac{ \det(V^{i}_{t})^{1/2}\det(\lambda I)^{-1/2}}{\delta}})\bigg)^{2} \label{eq:Conf_set_radius_unAtt}
\end{align}
After finding high-probability confidence sets for the unknown parameter, the core step is implementing Optimism in the Face of Uncertainty (OFU) principle. At any time $t$, we choose a parameter  $\tilde{\Theta}^{i}_t \in \mathcal{S}^i_c\cap \mathcal{C}^{i}_{t}(\delta)$ such that:

\begin{align}
J(\tilde{\Theta}^{i}_t, Q_*, R_*^i)\leq \inf\limits_{\Theta^{i} \in \mathcal{C}^{i}_t(\delta)\cap \mathcal{S}^i_c }J(\Theta^{i},Q_*,R^i_*)+\frac{1}{\sqrt{t}}. \label{eq:nonconvexOpt} \end{align}

Then, by using the chosen parameters as if they were the true parameters, the linear feedback gain $K(\tilde{\Theta}^{i}, Q_*, R_*^{i})$ is designed. We synthesized the control $\underline{u}^{i}_{t}=u^{i}_t+\nu_t(\mathcal{B}_{i})$  on (\ref{eq:dynam_by_theta3}) where $u^{i}_t=K(\tilde{\Theta}^{i}, Q_*, R_*^{i})x_t$. The extra exploratory noise $\nu_t \sim \mathcal{N}(\mu,\,\sigma_{\nu}^{2}I)\in \mathbb{R}^{d}$ with $\sigma_{\nu}^{2}=2\kappa^2\bar{\sigma}_{\omega}^2$ is the random ``more exploration" term. 

As can be seen in the regret bound analysis, recurrent switches in policy may worsen the performance, so a criterion is needed to prevent frequent policy switches. As such, at each time step $t$ the algorithm checks the condition $\det(V^{i}_{t})>2\det(V^{i}_{\tau})$ to determine whether updates to the control policy are needed where $\tau$ is the last time of policy update.

\subsubsection{Central Ellipsoid Construction}
Note that (\ref{eq:Conf_set_radius_unAtt}) holds regardless of the control signal $\underline{z}^{i}_t$. The formulation above also holds for any actuation mode, being mindful that the
dimension of the covariance matrix changes. Even while actuating in the IExp phase, by applying augmentation technique, we can build a confidence set (which we call the central ellipsoid) around the parameters of the full actuation mode  thanks to extra exploratory noise. For $t\leq T^{i}_c$, this simply can be carried out by rewriting (\ref{eq:dynam_by_theta3}) as follows:
\begin{align}
	x _{t+1} =\Theta _{*}^\top \bar{z}^i_{t}+\omega_{t+1}, \quad \bar{z}^i_t=\begin{pmatrix} x_t \\ \bar{u}^i_t \end{pmatrix} \label{eq:dynam_by_theta2} 
\end{align}
where $\bar{z}^i_t\in \mathbb{R}^{n+d}$ and $\bar{u}^i_t\in \mathbb{R}^d$ is constructed by augmentation as follows

\begin{align}
\bar{u}^i_t(\mathcal{B}_i)=u^i_t+\nu(\mathcal{B}_i),\quad \bar{u}^i_t(\mathcal{B}^c_i)=\nu(\mathcal{B}^c_i). \label{eq:controlAlg}
\end{align}

By this augmentation, we can construct the central ellipsoid 

\begin{align}
&	\nonumber\mathcal{C}_{t}(\delta) =\{\Theta^\top \in R^{n \times (n+d)}  \mid \operatorname{Tr}((\hat{\Theta}_{t}-\Theta)^\top V_{t}(\hat{\Theta}_{t}-\Theta))\\
& \nonumber\leq \beta_{t}(\delta)  \} \\
	& \beta_t(\delta) =(\sigma_{\omega}\sqrt{2n\log(\frac{\det(V_{t})^{1/2}\det(\lambda I)^{-1/2}}{\delta}})+\lambda^{1/2}s)^{2}. \label{eq:Conf_set_radius_centralElips} 
\end{align} 

which is an input to Algorithm 2 and used to compute IExp duration. 

\subsection{Main steps of Algorithm 2}

The main steps of Algorithm 2 are quite similar to those of Algorithm 1 with a minor difference in confidence set construction. Algorithm 2 receives $V_{T^{i^*}_c}$, $Z_{T^{i^*}_c}$, and $X_{T^{i^*}_c}$ from Algorithm 1, using which for  $t>T^{i^*}_c$ we have 
\begin{align*}
V_{t}&=V_{T^{i^*}_c} + \sum_{s=T^{i^*}_c+1}^{t-1} {z}_{s}{{z}_{s}}^\top\\
Z_tX_t&=Z_{T^{i^*}_c}X_{T^{i^*}_c}^\top + \sum_{s=T^{i^*}_c+1}^{t-1} {z}_{s}{{x}_{s}}^\top
\end{align*}
and the confidence set is easily constructed.

The following theorem summarizes boundedness of state norm when Algorithm 1 and 2 are deployed.


\begin{thm} \label{lemma:stabilization}
\begin{enumerate}
\item  The IExp algorithm keeps the states of the underlying system actuating in any mode $i$ bounded with the probability at least $1-\delta$ during initial exploration, i.e., 

\begin{align} \label{eq:upperbound_state}
\nonumber \|x_t\|&\leq \frac{1}{1-\Upsilon_i}\big(\frac{\eta_i}{\Upsilon_i}\big)^{n+d_i}\bigg[G_iZ_t^{\frac{n+d_i}{n+d_i+1}}\beta^i_t(\delta)^{\frac{1}{2(n+d_i+1)}}+\\
&\quad (\sigma_{\omega}\sqrt{2n\log\frac{nt}{\delta}}+\|s I\|\sigma_{\nu}\sqrt{2d_i\log\frac{d_it}{\delta}})\bigg]=:\alpha_t^i,
\end{align}
for all modes $i \in \{1,..., 2^d\} $. 

\item For $t>T^{i^*}_c+\frac{(n+d_{i^*})\log(n+d_{i^*})+\log c^{i^*}-\log \chi_s}{\log\frac{2}{1-\Upsilon}}:=T_{rc}$ we, with probability at least $1-\delta$, have $\|x_t\|\leq 2\chi_s$ where
\begin{align}
\chi_s:=\frac{2\sigma_{\omega}}{1-\Upsilon}\sqrt{2n\log\frac{n(T-T^{i^*}_c)}{\delta}}.
\end{align}
\end{enumerate}
\end{thm}

From parts (1) and (2) of Theorem  \ref{lemma:stabilization} we define the following \textit{good events}:
\begin{align}
F^i_{t}=\{\omega \in\Omega  \mid \forall s \leq T^{i}_c, \left\lVert x_{s}\right\rVert \leq \alpha^i_{t} \}.\label{eq:GoodEven_state_unat} 
\end{align}
and 
\begin{align}
F^{op,c}_{t}=\{\omega \in\Omega  \mid \forall\;\; T^{i^*}_c\leq s \leq t, \left\lVert x_{s}\right\rVert^2 \leq X_c^2 \}.\label{eq:GoodEven} 
\end{align}
in which
\begin{align}
X_c^2=\frac{32n\sigma^2_{\omega}(1+\kappa^2)}{(1-\Upsilon)^2}\log \frac{n(T-T_c)}{\delta}.\label{eq:upperBoundOptimisimphase} 
\end{align}
where both the events are used for regret bound analysis and the former one specifically is used to obtain best actuating mode for initial exploration.

\subsection{Determining the Optimal Mode for IExp}
\label{deterBest}
We still need to specify the best actuating mode $i^*$ for initial exploration along with its corresponding upperbound $X_t^{i^*}$. Theorem \ref{lemma:estimationerror_timeMinumum22} specifies $i^*$. First we need the following Lemma.

\begin{lem} \label{lemma:estimationerror_timeMinumum}
At the end of initial exploration, for any mode $\forall i\in \{1,..., 2^d\}$ the following inequality holds
\begin{align}
||\hat{\Theta}_{T^i_{\omega}}-\Theta_*||_2\leq \frac{\mu^i_c}{\sqrt{T^i_{\omega}}} \label{eq:stab_neighb}
\end{align}
where $\mu^i_c$ is given as follows

\begin{align}
\nonumber \mu^i_c:= & \frac{1}{\sigma_{\star}}
 \bigg(\sigma_{\omega}\sqrt{n(n+d)
\log\big(1+\frac{\mathcal{P}_c}{\lambda (n+d)}\big)+2n\log\frac{1}{\delta}}+\\
&\sqrt{\lambda}s\bigg)
 \label{eq:kappaDefControllable}
\end{align}
with,
\begin{align*}
\mathcal{P}_c&:={X_{T^i_{\omega}}^{i}}^2(1+2{\kappa^i}^2)T^i_{\omega}+
4T^i_{\omega}\sigma^2_{\nu}d_i\log (dT^i_{\omega}/\delta)
\end{align*}

in which $T^i_{\omega}$ stands for initial exploration duration of actuating in mode $i$. Furthermore, if we define
\begin{align}
T^i_c:=\frac{4(1+\kappa)^2\mu^{i2}_c}{(1-\Upsilon)^2}\label{eq:T_cDef}
\end{align}
then for $T^i_{\omega}>T^i_c$, $||\hat{\Theta}_{T^i_{\omega}}-\Theta_*||_2\leq \frac{1-\Upsilon}{2(1+\kappa)}$ holds with probability at least $1-2\delta$.
\end{lem}
The proof is provided in Appendix \ref{appendix}.

\begin{thm}\label{lemma:estimationerror_timeMinumum22}
Suppose Assumptions \ref{Assumption 1} and \ref{Assumption2} hold true. Then for a system actuating in the mode $i$ during initial exploration phase, the following results hold true
\begin{enumerate}

\item $I_{F^i_t} \max_{1\leq s\leq t} \|x_s\|\leq x_t$ 
where $I_{F^i_t}$ is indicator function of set $F^i_t$ and
\begin{align}
& x_t=Y_{i,t}^{n+d_i+1}\\
& \nonumber Y_{i,t}:=\max \big(e, \lambda (n+d^i)(e-1), \frac{-\bar{L}_i+\sqrt{\bar{L}_i^2+4\bar{K}_i}}{2\bar{K}_i}\big),
\end{align}
with
\begin{align*}
&\bar{L}^i=(\mathcal{D}^i_1+\mathcal{D}^i_2)\big(2n\sigma_{\omega}\log\frac{1}{\delta}+\sigma_{\omega}\sqrt{\lambda}s^i\big)\log t +\\
&\mathcal{D}^i_3
\sqrt{\log t/\delta}+
(\mathcal{D}^i_1+\mathcal{D}^i_2)n\sigma_{\omega}(n+d_i)\times \\ &\bigg(\log\frac{(n+d_i)\lambda+2{\mathcal{V}^i_t}^2}{(n+d_i)\lambda}t+\log\frac{(1+2{\kappa^i}^2)}{(n+d_i)\lambda}t\bigg)\log t\\
&\quad \bar{K}^i=2(\mathcal{D}^i_1+\mathcal{D}^i_2)n\sigma_{\omega}(n+d_i)(n+d_i+1)\log t.
\end{align*}
where 
	\begin{align*}
	 &\mathcal{D}^i_1:=\frac{4}{1-\Upsilon_i}\big(\frac{\eta_i}{\Upsilon_i}\big)^{n+d_i}\bar{G}_i (1+2{\kappa^i}^2)^{\frac{n+d_i}{2(n+d_i+1)}}\\
	  &  \mathcal{D}^i_2:=\frac{4}{1-\Upsilon_i}\big(\frac{\eta_i}{\Upsilon_i}\big)^{n+d_i}\bar{G}_i 2^{\frac{n+d_i}{2(n+d_i+1)}}\mathcal{V}_T^i,\\ &\mathcal{D}^i_3:=\frac{n\sqrt{2}}{1-\Upsilon_i}\big(\frac{\eta_i}{\Upsilon_i}\big)^{n+d_i}\sigma_{\omega}
	\end{align*}
	
in which $\mathcal{V}^i_t=\sigma_{\nu}\sqrt{2d_i\log d_it/\delta}$ holds with probability least $1-\delta/2$.

\item The best actuating mode $i^*$ for initial exploration is,
\begin{align}
\nonumber i^*&=argmin_{i\in \{1,...,2^{\mathbb{B}}\}} Y_{i,T^i_{\omega}}^{n+d_i+1}\\
& s.t\;\; \;\; T^i_{\omega}\geq T^i_c 
\label{optimiza}
\end{align}

\item The upper-bound of state norm of system actuating in the mode $i^*$ during initial exploration phase can be written as follows:
\begin{align}
\|x_t\|\leq c_c^{i^*}(n+d_{i^*})^{n+d_{i^*}}\label{simpleBound}
\end{align}
for some finite system parameter-dependent constant $c_c^{i^{*}}$.
\end{enumerate}
\end{thm}

\begin{rem}\label{Remark:Find Best Actuating mode}
While optimization problem (\ref{optimiza}) cannot be solved 
analytically because
$T^i_c$ itself depends 
on $x_{T^i_{\omega}}$, it can be 
determined using Algorithm 3.
\end{rem}

\begin{algorithm} 
\label{alg3}

	\caption{\small Find best actuating mode $i^*$ and its corresponding $T_c^{i^*}$\normalsize} \label{alg:FindMode}
		\begin{algorithmic}[1]
		\STATE \textbf{Inputs:}$\lambda,\,
		\kappa,\,S^i>0,$$\,\delta>0,$$\,, \sigma_{\omega},\, \vartheta_i,\, \eta_i,\, \Upsilon_i \, \forall i$
		\FOR {$\forall i\in \{1,...,2^{\mathbb{B}}\} $}
		\STATE $T^i_{itr}=1$
 		\FOR {$t=1,2,..., T^i_{itr} $}
		\STATE  compute $T_c^i$ by (\ref{eq:T_cDef}) 
		\IF {$t<T_c^i$} 
		\STATE $T^i_{itr}=T^i_{itr}+1$
		\ELSE 
		\STATE $T^i_{\omega}=t$
		\ENDIF
		\ENDFOR
		\ENDFOR
		\STATE Compute $x_{T^i_{\omega}}=Y_{i,T^i_{\omega}}^{n+d_i+1}$ $\forall i\in \{1,...,2^{\mathbb{B}}\}$
		\STATE Solve 
		$i^*=\arg \min_{i\in \mathcal{B}_*} x_{T^i_{\omega}}$.
		\STATE \textbf{Outputs:} $i^*$ and $T^{i^*}_{\omega}$
		\end{algorithmic}
	\end{algorithm}

\subsection{Regret Bound}
Recall (\ref{eq:Reg}), the regret for the proposed strategy (IExp+\\
SOFUA) can be defined as follows:

\begin{align}
	\mathcal{R}_T =\mathbb {E}\left[\sum_{t=1}^{T} ( x_{t}^\top Q_*x_{t} + {\bar{u}^{i(t)}_{t}}^\top R_* \bar{u}_{t}^{i(t)}-J_*(\Theta_*,Q_*,R_*))\right] \label{eq:RegBB} 
\end{align}

where $i(t)=i^*$ for $t\leq T^{i^*}_c$ and $i(t)=1$ for $t> T^{i^*}_c$.

An upper-bound for $\mathcal{R}_T$ is given by the following theorem which is the next core result of our analysis. 

\begin{thm} (Regret Bound of IExp+SOFUA) \label{lemma:RegretBoundControllable}
Under Assumptions \ref{Assumption 1} and \ref{Assumption2}, with probability at least $1-\delta$ the algorithm SOFUA together with additional exploration algorithm IExp which runs for $T^{i^*}_{c}$ time steps achieves regret of  $\mathcal{O}\big((n+d_{i^*})^{(n+d_{i^*})}T^{i^*}_{rc}\big)$ for $t\leq T^{i^*}_{c}$ and $\mathcal{O}\big(poly(n+d)\sqrt{T-T^{i^*}_{rc}}\big)$ for $t>T^{i^*}_{c}$ where $\mathcal{O}(.)$ absorbs logarithmic terms.
\end{thm}	

\section{Numerical Experiment}
\label{simulation}
In this section, we demonstrate on a practical example how the use of our algorithms successfully alleviates state explosion during the initial exploration phase. We consider a control system with drift and control matrices to be set as follows:

\begin{align*}
	A_*=\begin{pmatrix}
		1.04 & 0 & -0.27 \\
		0.52 & -0.81 & 0.83\\
		 0 & 0.04 & -0.90
	\end{pmatrix},\; B_*=\begin{pmatrix}
	0.61 &-0.29 & -0.47\\
	0.58 & 0.25 &  -0.5\\
	0 & -0.72 & 0.29
\end{pmatrix}.
\end{align*}

We choose the cost matrices as follows:

\begin{align*}
	Q_*=\begin{pmatrix}
	0.65 &-0.08& -0.14 \\
		-0.08 & 0.57 & 0.26\\
		-0.14& 0.26& 2.5
	\end{pmatrix},\;R_*=\begin{pmatrix}
		0.14 &0.04& 0.05 \\
		0.04 &0.24 &0.08\\
		0.05 &0.08 &0.2
	\end{pmatrix}.
\end{align*}
	
The Algorithm 3 outputs the exploration duration $T^{i^*}_c=50s$ and best actuating mode $i^*$ for initial exploration with corresponding control matrix $B_*^{i^*}$ and $R_*^{i^*}$
\begin{align*}
B_*^{i^*}=\begin{pmatrix}
	0.61 &-0.29\\
	0.58 & 0.25\\
	0 & -0.72
\end{pmatrix},\;\;\;\;\;\; R_*^{i^*}=\begin{pmatrix}
		0.14 &0.04  \\
		0.04 &0.24
	\end{pmatrix}.
\end{align*}	

\begin{figure}
	\centering
	\vspace{5pt}
		\hspace{4pt}
		\includegraphics[trim=4cm 6cm 6cm 8cm, scale=.4]{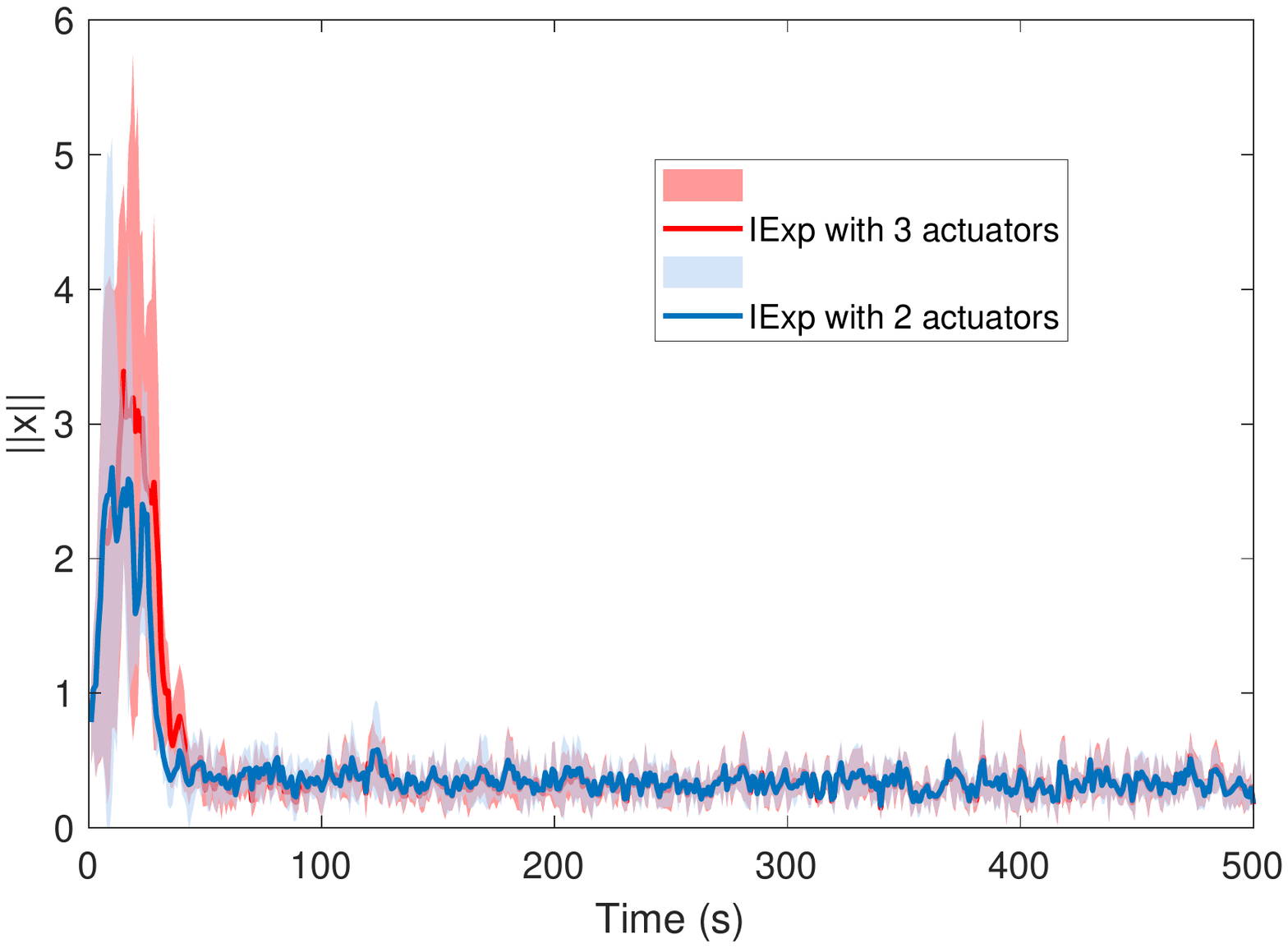}
	\hfill
	
		\hspace{4pt}
		\includegraphics[trim=4cm 7cm 6cm 8cm, scale=.4]{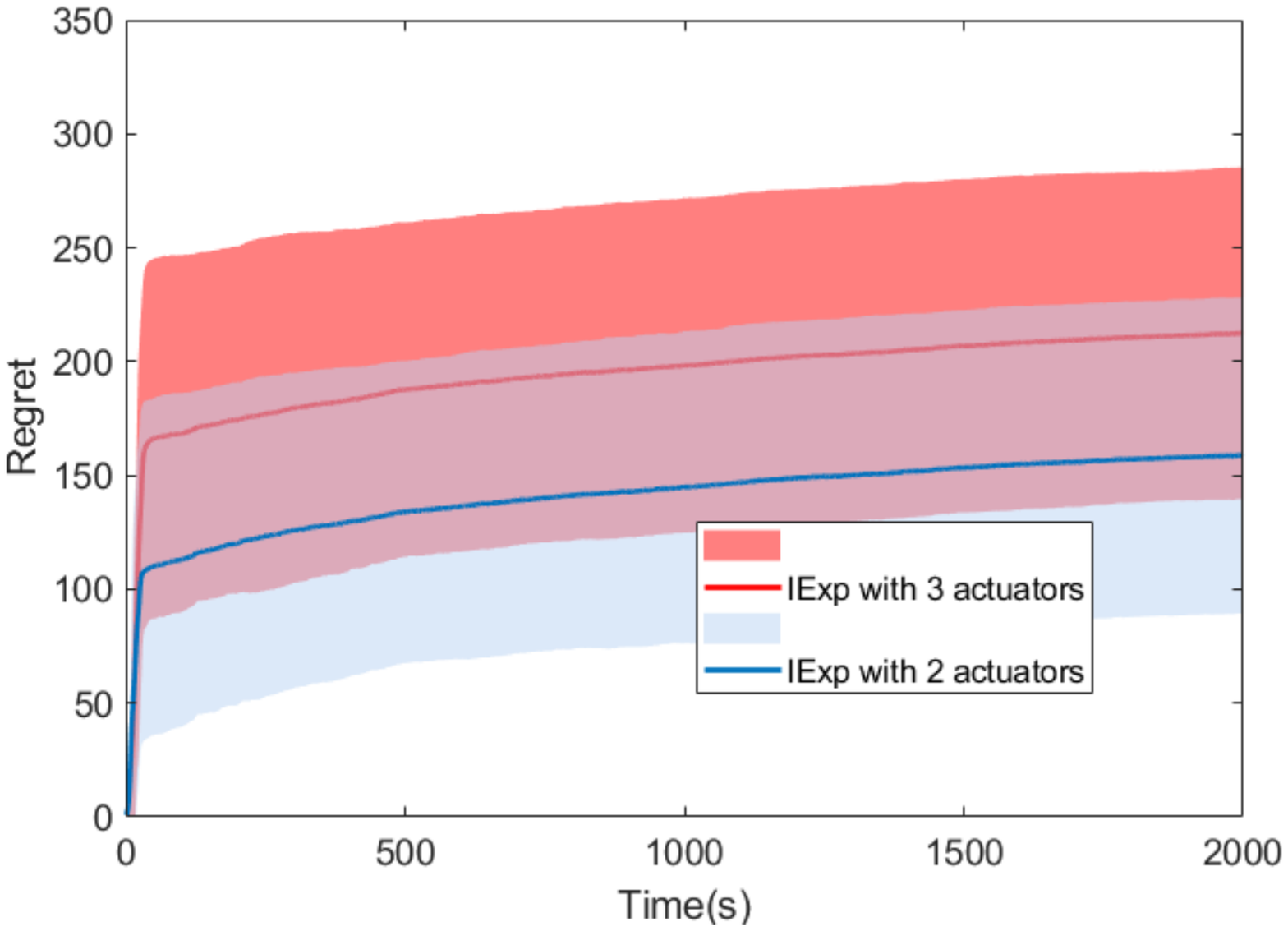}
	\hfill
	\caption{Top. State norm, Bottom. regret bound}
	\label{fig:naive_attacked_estimateANDregret}
\end{figure}

It has graphically been shown in \cite{abbasi2013online} that the optimization problem (\ref{eq:nonconvexOpt}) is generally non-convex for $n,d>1$. Because of this fact, we decided to solve optimization problem (\ref{eq:nonconvexOpt}) using a projected gradient descent method in Algorithm 1 and 2, with basic step

\begin{align}
\tilde{\Theta}^i_{t+1}\leftarrow PROJ_{\mathcal{C}^i_t(\delta)} \bigg(\tilde{\Theta}^i_{t}-\gamma  \nabla_{\Theta^i}(L^itr (P(\Theta^i,Q_*,R^{i}_*))) \bigg)
\end{align}

where $L^{i}=\bar{\sigma}^2_{\omega}+\vartheta^2\bar{\sigma}^2_{\nu}$ for $i=i^*$ and $L^{i}=\bar{\sigma}^2_{\omega}$ for $i=1$. $\nabla_{\Theta^i}f$ is the gradient of $f$ with respect to $\Theta^i$. $\mathcal{C}^i_t(\Theta^i)$ is the confidence set, $PROJ_g$ is Euclidean projection on $g$ and finally $\gamma$ is the step size. Computation of gradient $\nabla_{\Theta^i}$ as well as formulation of projection has been explicited in \cite{abbasi2013online}, similar to which we choose the learning rate as follows:
\begin{align*}
	\gamma=\sqrt{\frac{0.001}{tr(V^i_t)}}.
\end{align*}

We apply the gradient method for 100 iterations to solve each OFU optimization problem and apply the projection technique until the projected point lies
inside the confidence ellipsoid. The inputs to the OFU algorithm are $T=10000$, $\delta=1/T$, $\lambda=1$, $\sigma_{\omega}=0.1$, $s=1$ and we repeat simulation $10$ times. 

As can be seen in Fig. \ref{fig:naive_attacked_estimateANDregret}, the maximum value of the state norm (attained during the initial exploration phase) is smaller when using mode $i^*$ than when all actuators are in action.


Regret-bound for both cases is linear during initial exploration phase, however SOFUA guarantees $\mathcal{O}(\sqrt{T})$ regret for $T>50 s$.

\section{Conclusion}\label{conclusion}
In this work, we proposed an OFU principle-based controller for over-actuated systems, which combines
a step of ”more-exploration” (to produce a stabilizing
neighborhood of the true parameters while guaranteeing a bounded state during exploration) with one of ”optimism”, which efficiently
controls the system. Due to the redundancy, it is possible to further optimize the speed of convergence of the exploration phase to the stabilizing neighborhood by choosing over actuation modes, then to switch to full actuation to guarantee an  $\mathcal{O}(\sqrt{T})$ regret in closed-loop, with
polynomial dependency on the system dimension.

A natural extension of this work is to classes of systems in which some modes are only stabilizable. Speaking more broadly, the theme of this paper also opens the door to more applications of switching as a way to facilitate learning-based control of unknown systems, some of which are the subject of current work.


\bibliography{ifacconf}    

\begin{thebibliography}{13}
\providecommand{\natexlab}[1]{#1}
\providecommand{\url}[1]{\texttt{#1}}
\providecommand{\urlprefix}{URL }
\expandafter\ifx\csname urlstyle\endcsname\relax
  \providecommand{\doi}[1]{doi:\discretionary{}{}{}#1}\else
  \providecommand{\doi}{doi:\discretionary{}{}{}\begingroup
  \urlstyle{rm}\Url}\fi

\bibitem[{Abbasi and Szepesv{\'a}ri(2011)}]{abbasi2011regret}
Abbasi, Y. and Szepesv{\'a}ri, C. (2011).
\newblock Regret bounds for the adaptive control of linear quadratic systems.
\newblock In \emph{Proceedings of the 24th Annual Conference on Learning
  Theory}, 1--26.

\bibitem[{Abbasi-Yadkori(2013)}]{abbasi2013online}
Abbasi-Yadkori, Y. (2013).
\newblock Online learning for linearly parametrized control problems.
\newblock \emph{UAlberta}.

\bibitem[{Anderson and Moore(1971)}]{anderson1971linear}
Anderson, B.D. and Moore, J.B. (1971).
\newblock \emph{Linear Optimal Control [by] Brian DO Anderson [and] John B.
  Moore}.
\newblock Prentice-hall.

\bibitem[{Bertsekas(2011)}]{bertsekas2011dynamic}
Bertsekas, D.P. (2011).
\newblock Dynamic programming and optimal control 3rd edition, volume ii.
\newblock \emph{Belmont, MA: Athena Scientific}.

\bibitem[{Campi and Kumar(1998)}]{campi1998adaptive}
Campi, M.C. and Kumar, P. (1998).
\newblock Adaptive linear quadratic gaussian control: the cost-biased approach
  revisited.
\newblock \emph{SIAM Journal on Control and Optimization}, 36(6), 1890--1907.

\bibitem[{Chen and Hazan(2021)}]{chen2020black}
Chen, X. and Hazan, E. (2021).
\newblock Black-box control for linear dynamical systems.
\newblock In \emph{Conference on Learning Theory}, 1114--1143. PMLR.

\bibitem[{Cohen et~al.(2019)Cohen, Koren, and Mansour}]{cohen2019learning}
Cohen, A., Koren, T., and Mansour, Y. (2019).
\newblock Learning linear-quadratic regulators efficiently with only sqrt(t)
  regret.
\newblock \emph{International Conference on Machine Learning}, 1300--1309.

\bibitem[{Faradonbeh et~al.(2020)Faradonbeh, Tewari, and
  Michailidis}]{faradonbeh2017optimism}
Faradonbeh, M.K.S., Tewari, A., and Michailidis, G. (2020).
\newblock Optimism-based adaptive regulation of linear-quadratic systems.
\newblock \emph{IEEE Transactions on Automatic Control}, 66(4), 1802--1808.

\bibitem[{Ibrahimi et~al.(2012)Ibrahimi, Javanmard, and
  Roy}]{ibrahimi2012efficient}
Ibrahimi, M., Javanmard, A., and Roy, B.V. (2012).
\newblock Efficient reinforcement learning for high dimensional linear
  quadratic systems.
\newblock In \emph{Advances in Neural Information Processing Systems},
  2636--2644.

\bibitem[{Lale et~al.(2020{\natexlab{a}})Lale, Azizzadenesheli, Hassibi, and
  Anandkumar}]{lale2020explore}
Lale, S., Azizzadenesheli, K., Hassibi, B., and Anandkumar, A.
  (2020{\natexlab{a}}).
\newblock Explore more and improve regret in linear quadratic regulators.
\newblock \emph{arXiv preprint arXiv:2007.12291}.

\bibitem[{Lale et~al.(2020{\natexlab{b}})Lale, Azizzadenesheli, Hassibi, and
  Anandkumar}]{lale2020regret}
Lale, S., Azizzadenesheli, K., Hassibi, B., and Anandkumar, A.
  (2020{\natexlab{b}}).
\newblock Regret bound of adaptive control in linear quadratic gaussian (lqg)
  systems.
\newblock \emph{arXiv}.

\bibitem[{Lale et~al.(2021)Lale, Azizzadenesheli, Hassibi, and
  Anandkumar}]{lale2020adaptive}
Lale, S., Azizzadenesheli, K., Hassibi, B., and Anandkumar, A. (2021).
\newblock Adaptive control and regret minimization in linear quadratic gaussian
  (lqg) setting.
\newblock In \emph{2021 American Control Conference (ACC)}, 2517--2522. IEEE.

\bibitem[{Mania et~al.(2019)Mania, Tu, and Recht}]{mania2019certainty}
Mania, H., Tu, S., and Recht, B. (2019).
\newblock Certainty equivalence is efficient for linear quadratic control.
\newblock \emph{Advances in Neural Information Processing Systems}, 32.

\end{thebibliography}
\onecolumn
\section{Analysis}
 \label{sec:analysis}

In  this  section,  we  dig  further  and  provide  a numerical experiment and rigorous  analysis  of  the  algorithms,  properties  of  the  closed-loop  system, and regret bounds.  The most technical results, proofs and lemma can be found in the Appendix.

\subsection{Stabilization via IExp (proof of Theorem \ref{lemma:stabilization}. (1))}

This section attempts to upper-bound the state norm during initial exploration phase. We carry out this part regardless of which mode has been chosen for initial exploration. 

During the initial exploration the state recursion of system actuating in mode $i$ is written as follows:
\begin{align}
x_{t+1}=
\begin{cases}
\tilde{A}_t x_t+\tilde{B}^i_tu_t^i+B_*\nu_t+M^i_tz^i_t+\omega_t &\quad t \not\in\tau_{T_c^i} \\
A_*x_t+B_*\bar{u}^i_t+\omega_t &\quad t \in \tau_{T_c^i}\\
\end{cases}
\end{align}	
where $M^i_t=(\Theta^i_*-\tilde{\Theta}^i_t)$. The state update equation can be written as follows: 
\begin{align}
x_{t+1}=\Gamma^i_tx_t+r_t \label{eq:stateUpdateGen}
\end{align}
where 
\begin{align}
\Gamma^i_{t+1}=
\begin{cases}
\tilde{A}^i_t+\tilde{B}^i_t K(\tilde{\Theta}^i_t, Q_*, R_*^i) &\quad t \not\in\tau_{T_c^i} \\
A_*+B^i_*K(\tilde{\Theta}^i_t, Q_*, R_*^i) &\quad t \in \tau_{T_c^i}\\
\end{cases}
\end{align}	
and 
\begin{align}
r_{t+1}=
\begin{cases}
B_*\nu_t+M^i_tz^i_t+\omega_t &\quad t \not\in\tau_{T_c^i} \\
B_*\nu_t+\omega_{t} &\quad t \in \tau_{T_c^i}\\
\end{cases} \label{eq:Rs}
\end{align}	
By propagating the state back to time step zero, the state update equation can be written as:
\begin{align}
x_{t}=\prod_{s=0}^{t-1}\Gamma^i_s x_0+\sum_{k=1}^{t}\bigg(\prod_{s=k}^{t-1}\Gamma^i_s\bigg) r_k. \label{eq:Dyn_attp} 
\end{align}	
Recalling Assumptions \ref{Assumption2} we have

\begin{align}
\max_{t\leq T}||A_*+B^i_*K(\tilde{\Theta}^i_t, Q_*, R_*^i)||\leq \eta^i_c,\;\;\; \max_{t\leq T}\|\tilde{A}_t+\tilde{B}_t^iK(\tilde{\Theta}^i_t, Q_*, R_*^i)\|\leq \Upsilon_i<1.
\end{align}	

Now, by assuming $x_0=0$  it yields 
\begin{align} 
\nonumber \|x_t\|&\leq \big(\frac{\eta^i}{\Upsilon_i}\big)^{n+d_i}\sum_{k=1}^{t} \Upsilon_i^{t-k+1}\|r_k\|\leq \big(\frac{\eta^i}{\Upsilon_i}\big)^{n+d_i} \max_{1\leq k\leq t} \|r_k\|\sum_{k=1}^{t} \Upsilon_i^{t-k+1}\\
&=\frac{1}{1-\Upsilon_i}\big(\frac{\eta^i_c}{\Upsilon_i}\big)^{n+d_i} \max_{1\leq k\leq t} \|r_k\|.
\end{align}
On the other hand, we have 
\begin{align}
\|r_{k}\|\leq
\begin{cases}
\|M^i_kz^i_k\|+\|B_*\nu_k+\omega_k\| &\quad t \not\in\tau_{T_c^i} \\
\|B_*{\nu}_k+\omega_{k}\| &\quad t \in \tau_{T_c^i}\\
\end{cases} 
\end{align}
which results in 
\begin{align}
\max_{k\leq t}\|r_{k}\|\leq
\max_{k\leq t, t \notin \tau_{T^i_c}}\|M^i_kz^i_k\|+\max_{k\leq t}\|(sI){\nu}_k+\omega_k\|
\end{align}
where in second term from right hand side we applied the fact that $trace(\tilde{B}^\top \tilde{B})\leq s$ (see Assumption \ref{Assumption2}).

Now applying Lemma \ref{lemma:AbbsiYadkori_boundOnnoSometimes} and union bound argument one can write  
\small
\begin{align} \label{eq:upperbound_state}
\nonumber ||x_t||&\leq \frac{1}{1-\Upsilon_i}\big(\frac{\eta_i}{\Upsilon_i}\big)^{n+d_i}\bigg[G_iZ_t^{\frac{n+d_i}{n+d_i+1}}\beta^i_t(\delta)^{\frac{1}{2(n+d_i+1)}}+\\
&\quad (\sigma_{\omega}\sqrt{2n\log\frac{nt}{\delta}}+\|s I\|\sigma_{\nu}\sqrt{2d_i\log\frac{d_it}{\delta}})\bigg]=:\alpha_t^i, \; \forall i \in \{1,..., |\mathcal{B}^*_a|\} 
\end{align}
\normalsize
where $d_i$ stands for the number of actuators of an actuating mode $i$ and similarly any subscripts and superscripts $i$ denotes the actuating mode $i$.

The policy explicited in Algorithm 1 keeps the states of the underlying system bounded with the probability at least $1-\delta$ during initial exploration which is defined as the "good event" $F^i_t$
\begin{align}
F^i_{t}=\{\omega \in\Omega  \mid \forall s \leq T^{i}_c, \left\lVert x_{s}\right\rVert \leq \alpha^i_{t} \}.\label{eq:GoodEven_state_unat} 
\end{align}
A second "good event" is associated with the confidence set for an arbitraty mode $i$ defined as:
\begin{align}
E^i_{t}=\{\omega \in\Omega  \mid \forall s \leq t, \Theta_{*}^i \in \mathcal{C}^i_{s}(\delta/4)  \}\label{eq:GoodEven_set_unat} 
\end{align}
\subsection{Determining the exploration time and best mode for IExp} \label{secBestModeTime}
\subsubsection{Exploration duration}
Given the constructed central confidence set $\mathcal{C}_T$, we aim to specify the time duration $T^{i^*}_{c}$ that guarantees the parameter estimate resides within stabilizing neighborhood. For this, we need to lower bound the smallest eigenvalue of co-variance matrix $V_t$. The following lemma adapted from \cite{lale2020explore}, named persistence of excitation during the extra exploration, provides this lower bound.
\begin{lem}\label{Theorem:LowerBoundCovariance}
For the initial exploration period of $T_{\omega}\geq \frac{6n}{c_p^{\prime}}\log(12/\delta)$  we have
\begin{align}
\lambda_{\min}(V_{T_{\omega}})\geq \sigma_{\star}^2 T_{\omega} \label{eq:lowerBound}
\end{align}
with probability at least $1-\delta$ where $\sigma_{\star}^2=\frac{c_p^{\prime}\sigma_1^2}{16}$, $c_p^{\prime}:=\min\{c_p,c^{\prime \prime}_p\}$, and
\begin{align*}
c_p=\frac{\bar{\sigma}^2_{\omega}-\sigma^2_1-4\bar{\sigma}^2_{\nu}(1+\frac{\sigma^2_2}{2\bar{\sigma}^2_{\nu}})\exp(\frac{-\sigma_2^2}{\bar{\sigma}^2_{\nu}})}{\sigma_2^2}\\
c^{\prime \prime}_p=\frac{\frac{\bar{\sigma}^2_{\omega}}{2}-4\bar{\sigma}^2_{\nu}(1+\frac{\sigma^2_3}{2\bar{\sigma}^2_{\nu}})\exp(\frac{-\sigma_3^2}{\bar{\sigma}^2_{\nu}})}{\sigma_2^2}-0.5\bar{\sigma}^2_{\nu}\exp (\frac{-\sigma^2_3}{2\bar{\sigma}^2_{\nu}})
\end{align*}
for any $\sigma^2_1\leq \sigma^2_2$ and $\sigma^2_3$ such that $c_p, c^{\prime \prime}_p>0$.
\end{lem}

The following lemma gives an upper-bound for the parameter estimation error at the end of time $T$ which will be used to compute the minimum extra exploitation time $T_{\omega}$.
\begin{lem}\label{lemma:estimationerror}
Suppose assumption 1 and 2 holds. For $T\geq poly (\sigma_{\omega}^2,\sigma_{\nu}^2,n\log(1/\delta)$ having additional exploration leads to
\small
\begin{align}
||\hat{\Theta}_T-\Theta_*||_2\leq \frac{1}{\sigma_{\star}\sqrt{T}}\bigg(\sigma_{\omega}\sqrt{2n\log\big(\frac{n\det(V_T)^{1/2}}{\delta\det(\lambda I)^{1/2}}\big)}+\sqrt{\lambda}s\bigg)\label{eq:lemm5Statament}
\end{align}
\end{lem}
\normalsize
\begin{pf}
The proof is straightforward. First, a confidence set around the true but unknown parameters of the system is obtained which is given by  (\ref{eq:Conf_set_radius_unAtt}). Then applying (\ref{eq:lowerBound}) given by Lemma \ref{Theorem:LowerBoundCovariance} completes the proof.
\end{pf}

There is one more step to obtain the extra exploration duration, $T_{\omega}$ which is obtaining an upper-bound for the right hand side of  (\ref{eq:lemm5Statament}). Performing such a step allows us to state the following central result.

\subsubsection{Best Actuating mode for initial exploration}

Given the side information $\Upsilon_i$ and $\eta_i$s for all actuating modes $\forall i\in \{1,..., 2^{\mathbb{B}}\}$, using
the
bound (\ref{eq:upperbound_state}), we aim to find an actuating mode $i^*$ that provides the lowest
possible upperbound of state at first phase. This guarantees the state norm does not blow-up while minimizing the regret. The following lemma specifies the best mode $i^*$ to reach this goal. Theorem \ref{lemma:estimationerror_timeMinumum22} gives best actuating mode for IExp and its corresponding duration.

\subsection{Stabilization via SOFUA (proof of Theorem \ref{lemma:stabilization}. (2))}
After running the IExp algorithm for $t\leq T^{i^*}_c$ (or  $t\leq T^{i^*}_s$) noting that the confidence set is tight enough and we are in the stabilizing region, Algorithm 2 which leverages all the actuators comes into play. This algorithm has the central confidence set given by the Algorithm 1 as an input. The confidence ellipsoid for this phase is given as follows:
\begin{align}
	\nonumber\mathcal{C}_{t}(\delta) &=\{{\Theta}\in \mathbb{R}^{(n+d) \times n}  \mid\\
	&\quad \operatorname{Tr}((\hat{\Theta}_{t}-\Theta)^\top V_{t}(\hat{\Theta}_{t}-\Theta))\leq \beta_{t}(\delta)\} 
\end{align}
where 
\begin{align}
	\beta_t(\delta) =(\sigma_{\omega}\sqrt{2n\log(\frac{\det(V_{t})^{1/2}\det(\lambda I)^{-1/2}}{\delta}})+\lambda^{1/2}s)^{2} 
\end{align}
and 
\begin{align}
V_t=V_{T^{i^*}_e}+\sum_{T^{i^*}_e+1}^tz_tz_t^\top.
\end{align}
Now, we can define the good event $E_t$ for time $t>T^{i^*}_c$ 

\begin{align}
E_{t}=\{\omega \in\Omega  \mid \forall T^{i^*}_c\leq s \leq t, \Theta_* \in \mathcal{C}_{s}(\delta/4)  \}.\label{eq:GoodEven_set_unatFull} 
\end{align}

Now, we are ready to upperbound the state norm.

SOFUA keeps the states of the underlying system bounded with probability at least $1-\delta$. In this section, we define the "good event" $F^{op}_{t}$ for $t>T_c^{i^*}$.

Noting that for $t>T_c^{i^*}$ the algorithm stops applying the exploratory noise $\nu$, the state dynamics is written as follows: dynamics 
\begin{align}
 x_{t+1}=&\big(A_*+B_*K(\tilde{\Theta}_t, Q_*, R_*)\big)x_t+\omega_t
= M_t x_t+\omega_t
\label{eq:rDyn} 
\end{align}
where
\begin{align*}
M_t&=\big(A_*-\tilde{A}_{t-1}+\tilde{A}_{t-1}+B_*K(\tilde{\Theta}_t, Q_*, R_*)\\
&+\tilde{B}_{t}K(\tilde{\Theta}_{t},Q_*, R_*)-\tilde{B}_{t-1}K(\tilde{\Theta}_{t}, Q_*, R_*)\big).
\end{align*}
With controllability assumption for the $t>T^{i^*}_c$, if the event $E_t$ holds then $\|M_t\|< \frac{1+\Upsilon}{2}< 1$ for all $t\geq T^{i^*}_c$. Then starting from state $x_{T^{i^*}_c}$ one can write 
\begin{align}
\|x_t\|\leq \big(\frac{1+\Upsilon}{2}\big)^{t-T^{i^*}_c}\|x_{T^{i^*}_c}\|+\frac{2}{1-\Upsilon}\max_{T^{i^*}_c\leq s\leq t}\|\omega_s\|\label{eq:firstIneq}.
\end{align}
By applying union bound argument on the second term from right hand side of (\ref{eq:firstIneq}) and using the bound (\ref{simpleBound}), it is straight forward to show that
\begin{align*}
\|x_t\|&\leq \big(\frac{1+\Upsilon}{2}\big)^{t-T^{i^*}_c}c^{i^*}(n+d_{i^*})^{n+d_{i^*}}+\chi_s 
\end{align*}
where
\begin{align}
\chi_s:=\frac{2\sigma_{\omega}}{1-\Upsilon}\sqrt{2n\log\frac{n(T-T^{i^*}_c)}{\delta}}.
\end{align}
For $t>T^{i^*}_c+\frac{(n+d_{i^*})\log(n+d_{i^*})+\log c^{i^*}-\log \chi_s}{\log\frac{2}{1-\Upsilon}}:=T_{rc}$ we have $\|x_t\|\leq 2\chi_s$.
Now the "good event" $F^{op,c}_t$ is defined by
\begin{align}
F^{op,c}_{t}=\{\omega \in\Omega  \mid \forall\;\; T^{i^*}_c\leq s \leq t, \left\lVert x_{s}\right\rVert^2 \leq X_c^2 \}.\label{eq:GoodEven} 
\end{align}
in which
\begin{align}
X_c^2=\frac{32n\sigma^2_{\omega}(1+\kappa^2)}{(1-\Upsilon)^2}\log \frac{n(T-T_c)}{\delta}\label{eq:upperBoundOptimisimphase} 
\end{align}.

\subsection{Regret Bound Analysis}\label{regretBound}

\subsubsection{Regret decomposition}
From definition of regret, one can write

\begin{align}
\nonumber \mathcal{R}_T&=\sum_{t=1}^{T} (x_{t}^\top Q_*x_{t} + {\bar{u}^{i(t)}_{t}}^\top R_* \bar{u}_{t}^{i(t)})-TJ_*(\Theta_*,Q_*,R_*) \\
&
\nonumber =\sum_{t=0}^{T_{\omega}}\bigg(x_t^\top  Q x_t+{u^{i^*}_t}^\top R_*^{i^*}u^{i^*}_t+2{\nu_t}^\top R_*{\bar{u}^{i^*}_{t}}+{\nu_t}^\top R_*\nu_t\bigg)-TJ_*(\Theta_*,Q_*,R_*)\\
&+\sum_{t=T_{\omega}+1}^{T} (x_{t}^\top Q_*x_{t} + {\bar{u}_{t}}^\top R_* \bar{u}_{t})
\label{primalDecomp}
\end{align}
 Applying Bellman optimality equation (see \cite{bertsekas2011dynamic}) for LQ  systems actuating in any mode $i=i^*,1$, one can write
\begin{align*}
   &J(\tilde{\Theta}_{t-1}^i,Q_*, R_*^i)+ x_t^\top P(\tilde{\Theta}_{t-1}^i,Q_*, R_*^i)x_t\\ &=x_t^\top Q_*x_t+{u^i_t}^\top R_*^iu^i_t+\mathbb{E}\bigg[\big(\tilde{A}_{t-1}x_t+\tilde{B}^i_{t-1}u^i_t+\zeta_t\big)^\top P(\tilde{\Theta}_{t-1}^i,Q_*, R_*^i)\big(\tilde{A}_{t-1}x_t+\tilde{B}^i_{t-1}\underline{u}^i_t+\zeta_t\big)|\mathcal{F}_{t-1}\bigg]\\
    &=x_t^\top Q_*x_t+{u^i}_t^\top R_*^iu^i_t\\
  &+\mathbb{E}\big[\big(\tilde{A}_{t-1}x_t+\tilde{B}^i_{t-1}u^i_t\big)^\top P(\tilde{\Theta}_{t-1}^i,Q_*, R_*^i)\big(\tilde{A}_{t-1}x_t+\tilde{B}^i_{t-1}u^i_t\big)|\mathcal{F}_{t-1}\big]+\mathbb{E}\big[\zeta_t^\top P(\tilde{\Theta}_{t-1}^i,Q_*, R_*^i)\zeta_t|\mathcal{F}_{t-1}\big]\\
    &=x_t^\top Qx_t+{u^i}_t^\top R^iu^i_t+\mathbb{E}\big[\big(\tilde{A}_{t-1}x_t+\tilde{B}^i_{t-1}u^i_t\big)^\top P(\tilde{\Theta}_{t-1}^i,Q_*, R_*^i)\big(\tilde{A}_{t-1}x_t+\tilde{B}^i_{t-1}u^i_t\big)|\mathcal{F}_{t-1}\big]\\
    &+\mathbb{E}\bigg[x_{t+1}^\top P(\tilde{\Theta}_{t-1}^i,Q_*, R_*^i)x_{t+1}|\mathcal{F}_{t-1}\bigg]-\mathbb{E}\big[\big(A_*x_t+B^i_*u^i_t\big)^\top P(\tilde{\Theta}_{t-1}^i,Q_*, R_*^i)\big(A_*x_t+B^i_*u^i_t\big)| \mathcal{F}_{t-1}\big]\\
    &=x_t^\top Qx_t+{u^i_t}^\top R^iu^i_t+\mathbb{E}\bigg[x_{t+1}^\top P(\tilde{\Theta}_{t-1}^i,Q_*, R_*^i)x_{t+1}|\mathcal{F}_{t-1}\bigg]\\
    &+ \big(\tilde{A}_{t-1}x_t+\tilde{B}^i_{t-1}u^i_t\big)^\top P(\tilde{\Theta}_{t-1}^i,Q_*, R_*^i)\big(\tilde{A}_{t-1}x_t+\tilde{B}^i_{t-1}u^i_t\big)-\big(A_*x_t+B^i_*u^i_t\big)^\top P(\tilde{\Theta}_{t-1}^i,Q_*, R_*^i)\big(A_*x_t+B^i_*u^i_t\big)
\end{align*}

where $\zeta_t=B^i_*\nu_t(\mathcal{B}_i)+\omega_t$ for $t\leq T_{\omega}$ and $\zeta_t=\omega_t$ for $t> T_{\omega}$ and with slight abuse of notation $u_t^1=\bar{u}_t$. In third equality we applied the dynamics $x_{t+1}=A_*x_t+B^i_*u_t^i+\zeta$ and used the martingale property of process noise $\zeta_t$.

Now taking summation up to time  $T>T_{\omega}$ and redefining $J(\Theta^i, Q_*, R_*)= {\sigma_{\zeta} }^2Tr(P(\Theta^i, Q_*, R_*))$ gives

\begin{align}
&\sum_{t=0}^{T}\bigg(x_t^\top Q x_t+{u^{i(t)}_t}^\top R^{i(t)}{u}^{i(t)}_t\bigg)=\sum_{t=0}^{T}J(\tilde{\Theta}^{i(t)}_{t-1}, Q_*, R^{i(t)}_*)+R_1-R_2-R_3\\
&=\sum_{t=0}^{T_{\omega}}\bigg(\sigma^2_{\nu}Tr(P(\tilde{\Theta}_{t-1}^{i^*}, Q_*, R^{i^*}_*))B^{i^*}_*{B_*^{i^*}}^\top\bigg)\\
&+\sum_{t=0}^{T_{\omega}}\bigg(\bar{\sigma}^2_{\omega}Tr(P(\tilde{\Theta}_{t-1}^{i^*}, Q_*, R^{i^*}_*))\bigg)+\sum_{t=T_{\omega}+1}^{T}\bigg(\bar{\sigma}^2_{\omega}Tr(P(\tilde{\Theta}_{t-1}, Q_*, R_*))\bigg) +R_1-R_2-R_3 
 \label{decomb_toomuch}
\end{align}

where

\begin{align}
R_1=\sum_{t=0}^{T}\bigg(x^\top_tP(\tilde{\Theta}^{i(t)}_{t-1}, Q_*, R_*^{i(t)})x_t-\mathbb{E}\big[x^\top_{t+1}P(\tilde{\Theta}^{i(t)}_t, Q_*, R_*^{i(t)})x_{t+1}|\mathcal{F}_{t-1}\big]\bigg),
\end{align}

 \begin{align}
R_2=\sum _{t=0}^{T} \mathbb{E}[x_{t+1}^\top\big(P(\tilde{\Theta}_{t-1}^{i(t)}, Q_*, R_*^{i(t)})-P(\tilde{\Theta}_{t}^{i(t)}, Q_*, R_*^{i(t)})\big)x_{t+1}|\mathcal{F}_{t-1}]\label{eq:R2}  
\end{align}

and 

\begin{align}
\nonumber R_3 &=\sum _{t=0}^{T} \bigg((\tilde{A}_{t-1}x_{t}+\tilde{B}^{i(t)}_{t-1}{u}^{i(t)}_{t})^\top P^i(\tilde{\Theta}^i_{t-1})(\tilde{A}_{t-1}x_{t}+\tilde{B}^i_{t-1}{u}^{i(t)}_{t})\\
&\quad-(A_{*}x_{t}+B_{*}^{i(t)}{u}^{i(t)}_{t})^\top  P(\tilde{\Theta}^{i(t)}_{t-1}, Q_*, R_*^{i(t)})(A_{*}x_{t}+B^{i(t)}_{*}{u}^{i(t)}_{t})\bigg)\label{eq:R3}
\end{align}

with $i(t)=i^*$ for $t\leq T_{\omega}$,  and $i(t)=1$ when $t> T_{\omega}$ for which we drop the corresponding super/sub scripts abusively.

Recalling optimal average cost value formula and taking into account that the extra exploratory noise $\nu_t$ is independent of process noise $\omega_t$, for duration $t<T_{\omega}$ that system actuates in the mode $i^*$, the term $J(\tilde{\Theta}^{i^*}_{t-1},Q_*, R_*^{i^*})$ is decomposed as follows
\begin{align}
J(\tilde{\Theta}^{i^*}_{t-1},Q_*, R_*^{i^*})=\sigma_{\nu}^2Tr\big(P(\tilde{\Theta}^{i^*}_{t-1},Q_*, R_*^{i^*})B^{i^*}_*{B^{i^*}_*}^\top\big)+\bar{\sigma}_{\omega}^2Tr\big(P(\tilde{\Theta}^{i^*}_{t-1},Q_*, R_*^{i^*})\big).  
\end{align}

From given side information (\ref{Assum3_4}) (Assumption \ref{Assumption2}), one can write
\begin{align}
\nonumber&\bar{\sigma}_{\omega}^2Tr(P(\tilde{\Theta}^{i^*}_{t-1},Q_*, R_*^{i^*}))=J(\tilde{\Theta}^{i^*}_{t-1},Q_*, R_*^{i^*})\leq J(\tilde{\Theta}^{i^*}_{t-1},Q_*, R_*^{i^*})\\
&\leq J(\Theta_*,Q_*, R_*)+\gamma^{i^*} \label{important}
\end{align}
which results in
 \begin{align}
\sum_{t=0}^{T}\bigg(x_t^\top Q_*x_t+{{u}_t^{i(t)}}^\top R_*^{i(t)}{u}^{i(t)}_t\bigg)\leq \sigma_{\nu}^2T_{\omega}D||B^{i^*}_*||_F^2+ TJ_*(\Theta_*,\omega_t)+T_{\omega}\gamma^{i^*}+R_1-R_2-R_3
 \end{align}
 Combining (\ref{important}) with (\ref{decomb_toomuch}) and (\ref{primalDecomp}) for both controllable settings under the events $F^{i^*}_{T_c^{i^*}}\cap E_{T_c^{i^*}}$ for $t\leq T_c^{i^*}$ and $F^{opt}_T\cap E_T$ for $t\geq T_c^{i^*}$ (and for stabilizable setting with its corresponding events) the regret can be upper-bounded as follows:
\begin{align}
\mathcal{R}(T)&\leq \sigma_{\nu}^2T_{\omega}D||B^{i^*}_*||_F^2+\gamma^{i^*}T_{\omega}+R_0+ R_1-R_2-R_3 \label{eq:Decompos_reg}
\end{align}
 where
\small
\begin{align}
R_0=\sum_{t=0}^{T_{\omega}}\big(2{\nu_t}^\top R_*\underline{u}^{i^*}_t+{\nu_t}^\top R_*\nu_t\big).\label{eq:R_0}
\end{align}
\normalsize
 The term $R_0$ given by (\ref{eq:R_0}) which is direct effect of extra exploratory noise on the regret bound has same upper-bound for both controllable and stabilizable settings which is given by following lemma.

\begin{lem} (Bounding $R_0$) \label{lemma:R_0Bound}
On the event $E\cap F^{i^*}$, the term $R_0$ defined by (\ref{eq:R_0}) has the following upper bound:
\begin{align}
\nonumber|R_0|\leq d &\sigma_{\nu}\sqrt{B_{\delta}}+d\|R_*\|\sigma^2_{\nu}\\
&\times \big(T_{\omega}+\sqrt{T_{\omega}}\log\frac{4dT_{\omega}}{\delta}\sqrt{\log\frac{4}{\delta}}\big) 
\end{align}
where
\begin{align*}
\nonumber & B_{\delta}=8 \big(1+T_{\omega}\kappa^2\|R_*\|^2(n+d_{i^*})^{2(n+d_{i^*})}\big)\\
&\times\log \bigg(\frac{4d}{\delta}\big(1+T_{\omega}\kappa^2\|R_*\|^2(n+d_{i^*})^{2(n+d_{i^*})}\big)^{1/2}\bigg)
\end{align*}
\end{lem}
 The following subsequent sections gives rest of upperbounds. 

 \begin{lem} (Bounding $R_1$)\label{lemma:R_1Bound}
On the event $F^{i^*}_{T_c^{i^*}}\cap E^{i^*}_{T_c^{i^*}}$ for $t\leq T_c^{i^*}$ and $F^{c,op}_T\cap E^c_T$ for $t\geq T_c^{i^*}$, with probability at least $1-\delta/2$ for $t>T_{rc}$ the term $R_1$ is upper-bounded as follows:

\begin{align}
\nonumber R_1 &\leq k_{c,1} (n+d_{i^*})^{(n+d_{i^*})}(\sigma_{\omega}+| |B_*||\sigma_{\nu})\\ 
\nonumber &\quad\times n\sqrt{T_c^{i^*}}\log ((n+d^*)T_c^{i^*}/\delta)\\
\nonumber&\quad +k_{c,2}\sigma^2_{\omega}\frac{n\sqrt{n}}{(1-\Gamma)}\sqrt{t-T_c^{i^*}}\log \big(n(t-T_c^{i^*})/\delta\big)\\
\nonumber&\quad +k_{c,3}n\sigma^2_{\omega}\sqrt{t-T_{\omega}}\log (nt/\delta)\\
&\quad+k_{c,4}n(\sigma_{\omega}+||B_*||\sigma_{\nu})^2\sqrt{T_{\omega}}\log (nt/\delta)
\end{align}
for some problem dependent coefficients $k_{c,1},k_{c,2},k_{c,3},k_{c,4}$.
\end{lem}
 \begin{pf}
Proof follows as same steps as of \cite{lale2020explore}, with only difference that exploration phase is performed through actuating in the mode $i^*$ with corresponding number of actuators $d_{i^*}$.
\end{pf}

The following lemma upper-bounds the term $R_2$.
\begin{lem} (Bounding $|R_2|$) \label{lemma:R_2Bound}
On the event $F^{i^*}_{T_c^{i^*}}\cap E^{i^*}_{T_c^{i^*}}$ for $t\leq T_c^{i^*}$ and $F^{c,op}_T\cap E^c_T$ for $t\geq T_c^{i^*}$, it holds true that the term $R_2$ defined by (\ref{eq:R2}) is upper-bounded as
\small
\begin{align}
\nonumber &|R_2|\leq 2D{c^{i^* }}^2(n+d_{i^*})^{2(n+d_{i^*})}\\
\nonumber &\quad\times \big\{1+\log_2 \bigg(1 +\frac{{c^{i^*}}^2(1+2{\kappa^i}^2)(n+d_{i^*})^{2(n+d_{i^*})}T^{i^*}_c+4\sigma^2_{\nu}d_{i^*}\log\frac{8d_{i^*}T^{i^*}}{\delta}T^{i^*}}{\lambda} \bigg)^{n+d_{i^*}}\big\}\\
 &\quad+2D\frac{32n\sigma^2_{\omega}(1+\kappa^2)}{(1-\Upsilon)^2} \log\frac{n(T-T_c^{i^*})}{\delta} 
\log_2 \big(\frac{\bar{\lambda}}{\sigma_{\star}^2 (T_c^{i^*}+1)}\big)^{n+d}
\end{align}
\normalsize
 in which 
\begin{align*}
\nonumber \bar{\lambda}&:=\bar{\lambda}:=\lambda + {c^{i^*}}^2T^{i^*}_c(n+d_{i^*})^{2(n+d_{i^*})}(1+2{\kappa^i}^2)\\
\nonumber&\quad+4\sigma^2_{\nu}d_{i^*}\log(d_{i^*}T^{i^*}_c/ \delta)\\
&\quad+(T-T^{i^*}_c)\frac{32n\sigma^2_{\omega}(1+\kappa^2)}{(1-\Upsilon)^2} \log\frac{n(T-T^{i^*}_c)}{\delta}.
\end{align*}
\end{lem}
 
 The proof can be found in Appendix \ref{appendix} 

\begin{lem} (Bounding $|R_3|$) \label{lemma:R_3Bound}
On the event $F^{i^*}_{T_c^{i^*}}\cap E^{i^*}_{T_c^{i^*}}$ for $t\leq T_c^{i^*}$ and $F^{c,op}_T\cap E^c_T$ for $t\geq T_c^{i^*}$, the term $R_3$ defined by (\ref{eq:R3}) has the following upper bound:
\begin{align}
|R_3|=\mathcal{O}\big((n+d_{i^*})^{(2(n+d_{i^*}))}T_c^{i^*}+(n+d)n^2\sqrt{T-T_c^{i^*}}\big)
\end{align}
\end{lem}

Putting Everything Together, gives the overall regret bound which holds with probability at least $1-\delta$. This bound has been summarized by Theorem \ref{lemma:RegretBoundControllable}.



 \newpage
\section{Appendix} \label{appendix}

 \subsection{Technical Theorems and Lemmas}
\begin{lem}(Norm of Sub-gaussian vector)
For an entry-wise $R-$subgaussian vector $y\in \mathbb{R}^m$ the following upper-bound holds with probability at least $1-\delta$ 
\begin{align*}
    \|y\|\leq R\sqrt{2m\log(\frac{m}{\delta})}
\end{align*}
\end{lem}

\begin{lem} \label{Self_normalized_Bound} (Self-normalized bound for vector-valued martingales \cite{abbasi2011regret})
	 Let $F_k$ be a filtration, $z_k$ be a stochastic process adapted to $F_k$ and $\omega^i_k$ (where $\omega^i_k$ is the $i-$th element of noise vector $\omega_k$) be a real-valued martingale difference, again adapted to filtration $F_k$ which satisfies the conditionally sub-Gaussianity assumption (Assumption 2.4) with known constant $\sigma_{\omega}$. 
	 Consider the martingale and co-variance matrices:
	\begin{align*}
S^i_t:=\sum _{k=1}^{t} z_{k-1}\omega^i_k, \quad V_{t}=\lambda I+\frac{1}{\beta}\sum _{s=1}^{t-1}z_{t} z_{t}^T
	\end{align*}
	
then with probability of at least $1-\delta$, $0<\delta<1$ we have,
\begin{align}
\left\lVert S^i_t\right\rVert^2_{V_{t}^{-1}} \leq 2 \sigma_{\omega}^2\log \bigg(\frac{\det(V_t)^{1/2} }{\delta \det(\lambda I)^{1/2}}\bigg)
\end{align}	
\end{lem}
Given the fact that for controllable systems solving DARE gives a unique stabilizing controller, in this section we go through an important result from literature that show there is a strongly stabilizing neigborhood around the parameters of a system. This means that solving DARE for any parameter value in this neighborhood gives a controller which stabilizes the system with true parameters. 

\begin{lem}(Mania et al. \cite{mania2019certainty}) \label{Lemma:Mania}
There exists explicit constants $C_0$ and 
\begin{align*}
    \epsilon=poly(\underline{\alpha}^{-1}, \bar{\alpha}, \|A_*\|, \|B_*\|, \bar{\sigma}_{\omega},D,n,d)
\end{align*}
in which $\underline{\alpha}I\leq Q\leq \bar{\alpha}I$ and $\underline{\alpha}I\leq R\leq \bar{\alpha}I$ such that for any $\Theta^{\prime}\in\{\mathbb{R}^{n\times(n+d)} \;|\; \|\Theta^{\prime}-\Theta_*\|\leq \varepsilon,\; 0\leq \varepsilon\leq \epsilon\}$, we have 
\begin{align}
   J(K(\Theta^{\prime}), \Theta_*, Q,R)-J^*\leq C_0\varepsilon^2
\end{align}
where $J(K(\Theta^{\prime}), \Theta_*, Q,R)$ is infinite horizon performance of policy $K(\Theta^{\prime})$ applied on $\Theta_*$.
\end{lem}
 Lemma \ref{Lemma:Mania} implicitly says that for any estimates residing within stabilizing neighborhood, the designed controller, applied on true system, is stabilizing. 


\subsection{Confidence Set Construction}
The following theorem gives the confidence set for initial exploration phase and actuating mode $i$. Central confidence set and confidence set of actuating mode 1 can be similarly constructed.
\begin{thm}\label{thm:Conficence_Set_Attacked} (System Identification)
	Consider linear dynamics model (\ref{eq:dynam_by_theta3}) where $\omega_t$ and $\nu_t$ are independent random vectors, both satisfying Assumption \ref {Assumption 1} with a known $\sigma_{\omega}$ and $\sigma_{\nu}$. Let $trace({\Theta^i}^\top \Theta^i)\leq s$ (which is part of Assumptions \ref{Assumption2}) hold and $\hat{\Theta}^i_t$ be $l_2-$ regularized least square estimation at time $t$.
	 Then with probability at least $1-\delta$ we have $\Theta^i_*\in \mathcal{C}_t^i(\delta)$ where
\begin{align}
	\nonumber\mathcal{C}^{i}_{t}(\delta)& =\{{\Theta^i}^\top \in R^{n \times (n+d_i)}  \mid \\  & \nonumber\operatorname{Tr}((\hat{\Theta}^{i}_{t}-\Theta^{i})^\top V_{t}^i(\hat{\Theta}^{i}_{t}-\Theta^{i}))\leq \beta^{i}_{t}(\delta)  \}, \\
	\nonumber\beta^{i}_t(\delta) &=\bigg(\lambda^{1/2}s^i+\sigma_{\omega}\sqrt{2n\log(n\frac{\det(V^{i}_{t})^{1/2}\det(\lambda I)^{-1/2}}{\delta}})\\
	&+\|\bar{B}_*^i\|\sigma_{\nu}\sqrt{2d_i\log(d_i\frac{ \det(V^{i}_{t})^{1/2}\det(\lambda I)^{-1/2}}{\delta}})\bigg)^{2} 
\end{align}
\end{thm}
\begin{pf}

  From (\ref{eq:LSE_Solf}) we have:
  	\begin{align}
	\hat{\Theta}^i_t &=\operatorname*{argmin_{\Theta^{i}}} e(\Theta^{i})=({\underline{Z}_t^{i}}^\top \underline{Z}_t^{i}+\lambda I)^{-1}{\underline{Z}_t^{i}}^\top X_t
\end{align}
	where $\underline{Z}_t^{i}$ and $X_t$ are matrices whose rows are ${\underline{z}^{i}_{0}}^\top,..., {\underline{z}^{i}_{t-1}}^\top$ and $x_{1}^\top,...,x_{t}^\top$, respectively.
  and on the other hand considering the definition of $X_t$, $\underline{Z}_t^{i}$ and $W_t$ whose rows are $(\bar{B}^i_*\bar{\nu}^i_0+\omega_{0})^\top$,..., $(\bar{B}^i_*\bar{\nu}^i_{t-1}+\omega_{t-1})^\top$ the dynamic of system can be written as
  \begin{align*}
      X_t=\underline{Z}_t^{i}\Theta_*^i+W_t
  \end{align*}
  which leads to

  \begin{align}
  \nonumber\hat{\Theta}^i_{t} &=({\underline{Z}^i_t}^\top{\underline{Z}^i_t}+\lambda I)^{-1}{\underline{Z}^i_t}^\top(\underline{Z}^i_t \Theta _{*}^i+W_t)\\
  \nonumber&\quad =({\underline{Z}^i_t}^\top{\underline{Z}^i_t}+\lambda I)^{-1}({\underline{Z}^i_t}^\top \underline{Z}^i_t+\lambda I) \Theta^i_*-\lambda({\underline{Z}^i_t}^\top{\underline{Z}^i_t}+\lambda I)^{-1} \Theta^i_*\\
  \nonumber&\quad+({\underline{Z}^i_t}^\top{\underline{Z}^i_t}+\lambda I)^{-1}{\underline{Z}^i_t}^\top W_t \label{eq:tpcd}
  \end{align}
 Noting definition $V^i_{t}={\underline{Z}^i_t}^\top{\underline{Z}}^i_t+\lambda I$
  it yields
  \begin{align}
  \hat{\Theta}^i_{t}-\Theta^i_{*} &={{V}^i_{t}}^{-1}{\underline{Z}_t^i}^\top W_t+
  {V^i_t}^{-1}\lambda\Theta^i_{*}.
  \end{align}
  For an arbitrary random covariate $z^i$ we have,
  
  \begin{align}
  {z^i}^\top\hat{\Theta}^i_{t}-{z^i}^\top\Theta^i_{*} &=\langle\,z^i,{\underline{Z}^i_t}^\top W_t\rangle_{{V^i_{t}}^{-1}}+
  \langle\,z^i,\lambda\Theta^i_{*}\rangle_{{V^i_{t}}^{-1}}.
  \end{align}
  By taking norm on both sides one can write,
  \begin{align}
  \| {z^i}^\top\hat{\Theta}^i_{t}-{z^i}^\top\Theta^i_{*}\| &\leq \| z^i\|_{{V^i_{t}}^{-1}}\Bigg (\| {\underline{Z}}^\top_tW_t \|_{{V^i_{t}}^{-1}}+\|\lambda \Theta^i_*\|_{\bar{V}_{t}^{-1}} \Bigg) \leq \| z^i\|_{{V^i_{t}}^{-1}}\Bigg (\| {\underline{Z}}^\top_tW_t \|_{{V^i_{t}}^{-1}}+\sqrt{\lambda}s^i \Bigg).\label{eq:tobeFoll}
  \end{align}
  where in last inequality we applied the $\|\Theta^i_*\|^2\leq {s^i}^2$ (from Assumptions \ref{Assumption2} and \ref{Assumption3}) and teh fact that ${V^i_t}^{-1}\leq 1/\lambda$.
  
  Using Lemma \ref{Self_normalized_Bound}, $\| {\underline{Z}^i_t}^\top W_t \|_{{V^i_t}^{-1}}$ is bounded from above as
  \begin{align}
  \| {\underline{Z}^i_t}^\top W_t \|_{{V^i_t}^{-1}} &\leq \sigma_{\omega}\sqrt{2n\log(\frac{n\det(V^i_{t})^{1/2}\det(\lambda I)^{-1/2}}{\delta}})+ \|\bar{B}_*^i\|\sigma_{\nu}\sqrt{2d_i\log(\frac{d_i\det(V^i_{t})^{1/2}\det(\lambda I)^{-1/2}}{\delta}}).
  \end{align}
  
 By arbitrarily choosing  $z^i=V^i_{t}(\hat{\Theta}_{t}^i-\Theta^i_{*})$ and plugging it into (\ref{eq:tobeFoll}) it yields
  \begin{align}
  &\nonumber\| \hat{\Theta}^i_{t}-\Theta^i_{*}\ \|^2_{V^i_{t}}  \leq \| {V}^i_{t}(\hat{\Theta}_{t}^i-\Theta^i_{*})\|_{{V^i_{t}}^{-1}}\\
    \nonumber&\quad\Bigg (\sigma_{\omega}\sqrt{2n\log(\frac{\det({V}^i_{t})^{1/2}\det(\lambda I_{(n+d_i)\times (n+d_i)})^{-1/2}}{\delta}})+\|\bar{B}_*^i\|\sigma_{\nu}\sqrt{2d_i\log(\frac{d_i\det({V}^i_{t})^{1/2}\det(\lambda I_{(n+d_i)\times (n+d_i)})^{-1/2}}{\delta}})+\sqrt{\lambda}s^i\Bigg)
  \end{align}
  and since $\| \hat{\Theta}^i_{t}-\Theta^i_{*}\|_{V^i_{t}}= \|{V}^i_{t}(\hat{\Theta}^i_{t}-\Theta^i_{*})\|_{{V^i_{t}}^{-1}}$, the statement of Theorem \ref{thm:Conficence_Set_Attacked} holds true.
  	\end{pf}
\begin{lem} (\cite{abbasi2011regret})\label{lemma:AbbsiYadkori_boundOnnoSometimes}
For any $0\leq t\leq T$ and $\forall \in \{1,..., 2^{\mathbb{B}}\}$  we have that
\begin{align}
\max_{s\leq t, s\notin \tau_t}||(\Theta^i_*-\hat{\Theta}^i_s)^\top z^i_s||\leq GZ^{\frac{n+d_i}{n+d_i+1}}\beta^i_t(\delta/4)^{\frac{1}{2(n+d_i+1)}}
\end{align}

where $\tau_t$ is a set of finite number of times, with maximum cardinality $n+d_i$, occurring between any time interval $[0\;t]$ such that
$||(\Theta^i_*-\hat{\Theta}^i_s)^\top z^i_s||$ is not well controlled in those instances (see Lemma 17 and 18 of \cite{abbasi2011regret} for more details). 
\end{lem}
During initial exploration, while the system actuate in an arbitrary mode $i$ it constructs the central ellipsoid by using augmentation technique. The following lemma gives an upper-bound for the determinant of co-variance matrix of mode $i$, $V^i$ and that of central ellipsoid $V$ (for central ellipsoid $i=1$).  

\begin{lem}
Let the system actuate in an arbitrary mode $i$ and applies extra exploratory noise $\nu$. Further assume that the central ellipsoid is constructed by applying augmentation technique. Then with probability at least $1-\delta$ the determinant of the co-variance matrix $V^i$ ($i=1$ denotes the central ellipsoid) is given by
		\small
		\begin{align}
		\nonumber &\frac{\det(V^i_t)}{\det(\lambda I_{(n+d_i)\times (n+d_i)})}\leq\\
		& \Bigg(\frac{(n+d_i)\lambda+(1+2{\kappa^i}^2) {x_t}^2t+2{\mathcal{V}^i_t}^2t} {(n+d_i)\lambda}\Bigg)^{n+d_i}
	\end{align}
	\normalsize
where $\mathcal{V}^i_t=\sigma_{\nu}\sqrt{2d_i\log d_it/\delta}$.
\end{lem}
\begin{pf}
we can write 
	\begin{align}
		\nonumber det(V^i_t)&\leq \prod_{j=1}^{n+d_i}\Big(\lambda+\sum_{k=1}^{t-1}{z^i_{j}}_k^2\Big)\\ \nonumber&\quad\leq \bigg(\frac{\sum_{j=1}^{n+d_i}\big(\lambda+\sum_{k=1}^{t-1}{z^i_{j}}_k^2\big)}{n+d}\bigg)^{n+d_i}\\ \nonumber&\quad\leq\Bigg(\frac{(n+d_i)\lambda+ \sum_{k=1}^{t-1}\|x_k\|^2+2\|u^i_k\|+2\|\nu(\mathcal{B}_i)_k\|^2}{n+d_i}\Bigg)^{n+d_i}
	\end{align}	
		In second inequality, we applied AM-GM inequality and in the third inequality we apply the property $(a+b)^2\leq 2a^2+2b^2$. Furthermore, $\| u^i_k\|^2\leq {\kappa^i}^2 \|x_k\|^2$.
		Given $\max_{0\leq k\leq t}||\nu_k||=\mathcal{V}^i_t$ one can write: 
		\small
		\begin{align}
		\nonumber &\frac{\det(V^i_t)}{\det(\lambda I_{(n+d_i)\times (n+d_i)})}\leq\\
		& \Bigg(\frac{(n+d_i)\lambda+(1+2{\kappa^i}^2) {x_t}^2t+2{\mathcal{V}^i_t}^2t} {(n+d_i)\lambda}\Bigg)^{n+d_i}
	\end{align}
	\normalsize
where $\mathcal{V}^i_t=\sigma_{\nu}\sqrt{2d_i\log d_it/\delta}$ holds with probability least $1-\delta/2$. 
This completes the proof of (\ref{eq:stab_neighb}). Proof of the second statement of lemma is given in \cite{lale2020explore}.
\end{pf}
\subsection {Proofs of Lemma \ref{lemma:estimationerror_timeMinumum}\label{prooflemma:estimationerror_timeMinumum} and Theorem \ref{lemma:estimationerror_timeMinumum22} \label{prooflemma:estimationerror_timeMinumum22}}
\begin{pf}(Proof of Lemma \ref{lemma:estimationerror_timeMinumum})\label{prooflemma:estimationerror_timeMinumum}.
The proof directly follows by plaguing the upper-bound of $det(V_t)$ into Lemma \ref{lemma:estimationerror}.  
	\begin{align}
		\nonumber det(V_t)&\leq \prod_{i=1}^{n+d}\Big(\lambda+\sum_{k=1}^{t-1}z_{ki}^2\Big)\\ \nonumber&\quad\leq \bigg(\frac{\sum_{i=1}^{n+d}\big(\lambda+\sum_{k=1}^{t-1}z_{ki}^2\big)}{n+d}\bigg)^{n+d}\\ \nonumber&\quad=\Bigg(\frac{(n+d)\lambda+ \sum_{k=1}^{t-1}|| \bar{z}^{i^*}_k||^2+2||\zeta_k||^2}{n+d}\Bigg)^{n+d}
	\end{align}	
		In second inequality we applied AM-GM inequality and in the third inequality we apply the property $(a+b)^2\leq 2a^2+2b^2$. Furthermore, $|| \bar{z}^{i^*}_k||^2\leq (1+2\kappa^2)X_e^2$.
		Given $||\zeta_t||\leq 2\sigma_{\nu}\sqrt{d\log(4nt(t+1)/ \delta) }$ which holds with probability at least $1-\delta/2$ one can write: 
		
	\begin{align}
		\nonumber &\frac{\det(V_t)}{\det(\lambda I_{(n+d)\times (n+d)})}\leq\\
		& \Bigg(\frac{(n+d)\lambda+c^{i^*}(n+d^*)^{2(n+d^*)}(1+2\kappa^2)t+4\sigma^2_{\nu}(d-d^*)\log (8nt(t+1)/\delta)} {(n+d)\lambda}\Bigg)^{n+d}
	\end{align}
	
This completes the proof of (\ref{eq:stab_neighb}). Proof of the second statement of lemma is given in \cite{lale2020explore}.
\end{pf}

\begin{pf} (Proof of Theorem \ref{lemma:estimationerror_timeMinumum22}) \label{prooflemma:estimationerror_timeMinumum22}
Given (\ref{eq:upperbound_state}), we first upper-bound the term $G_i Z_T^{\frac{n+d_i}{n+d_i+1}}\beta^i_t(\delta/4)^{\frac{1}{2(n+d_i+1)}}$.
First, we can write
\begin{align}
\nonumber&{Z^i_T}^2=\max_{0\leq t\leq T}\|z_t\|^2\leq \max_{0\leq t\leq T}\|x_t\|^2+\max_{0\leq t\leq T}\|u^i_t+\nu_t(\mathcal{B}_i)\|^2\leq \max_{0\leq t\leq T}\|x_t\|^2+2 \|u^i_t\|^2+2\max_{0\leq t\leq T}\|\nu_t(\mathcal{B}_i)\|^2 \leq\\
&(1+2{\kappa^i}^2)\max_{0\leq t\leq T}\|x_t\|^2+2\max_{0\leq t\leq T}\|\nu_t(\mathcal{B}_i)\|^2\label{eq:boundzn}
\end{align}
which results in
\begin{align}
{Z^i_T}=\max_{0\leq t\leq T}||z^i_t||\leq \sqrt{(1+2{\kappa^i}^2)}x_T+\sqrt{2}\mathcal{V}_T^i
\end{align}

where $x_T:=\max_{0\leq t\leq T} ||x_t||$ and $\mathcal{V}_T^i:=\max_{0\leq t\leq T} ||\nu_t(\mathcal{B}_i)||$. On the other hand, given the definition of $\beta^i_t(\delta/4)$ one can write:
\begin{align}
\beta^i_t(\delta/4)^{\frac{1}{2(n+d_i+1)}}\leq \beta^i_t(\delta/4)^{\frac{1}{2}}\leq 4\sqrt{\beta^i_t(\delta)}
\end{align}
Combining the results give:
\begin{align}
G_i {Z^i_T}^{\frac{n+d_i}{n+d_i+1}}\beta^i_t(\delta/4)^{\frac{1}{2(n+d_i+1)}} &\leq 4G_i \sqrt{\beta^i_t(\delta)} \big(\sqrt{1+2{\kappa^i}^2}x_T+\sqrt{2}\mathcal{V}_T^i\big)^{\frac{n+d_i}{n+d_i+1}}\\ \quad & 4G_i \sqrt{\beta^i_t(\delta)}\big((1+2{\kappa^i}^2)^{\frac{n+d_i}{2(n+d_i+1)}}{x_T}^{\frac{n+d_i}{n+d_i+1}}+2^{\frac{n+d_i}{2(n+d_i+1)}}{\mathcal{V}_T^i}^{\frac{n+d_i}{n+d_i+1}}\big)
\end{align}
We also have
\begin{align*}
G_i=2 \bigg(\frac{2S(n+d_i)^{\frac{n+d_i+1}{2}}}{U^{0.5}}\bigg)^{1/(n+d_i+1)}
\end{align*}
where
\begin{align*}
U^i=\frac{U_0}{H}\;\;\;\; and \;\;\; U^i_0=\frac{1}{16^{n+d_i-2}(1\vee S^{2(n+d_i-2)})} 
\end{align*}
One can simply rewrite $G_i$ as follows:
\begin{align*}
G_i=\mathcal{C}H^{\frac{1}{2(n+d_i+1)}}
\end{align*}
where 
\small
\begin{align*}
\mathcal{C}:=2 \bigg(2S(n+d_i) \big(16^{n+d_i-2}(1\vee S^{2(n+d_i-2)}) \big)^{1/2}\bigg)^{1/(n+d_i+1)}
\end{align*}
\normalsize
and
\begin{align*}
H_i>\bigg(16\vee\frac{4S^2M_i^2}{(n+d_i)U^i_0}\bigg)
\end{align*}
with $M_i$ to be defined as follows
	\begin{align*}
		M_i=\sup_{Y> 0} \frac{\|\bar{B}_*^i\|\sigma_{\nu}\sqrt{n(n+d_i)\log\big(\frac{1+TY/\lambda(n+d_i)}{\delta}\big)}+\sigma_{\omega}\sqrt{n(n+d_i)\log\big(\frac{1+TY/\lambda(n+d_i)}{\delta}\big)}+\lambda^{1/2}s^i}{Y}
	\end{align*}
	Given the fact that $Y=\sup_{0\leq t\leq T} ||z_t||$. Then by having nonzero initial state $x_0$ then by defining $Y^*= \sqrt{(1+2{\kappa^i}^2)}\|x_0\|+\sqrt{2}\mathcal{V}_T^i$ we have 
	the following upper-bound for $M$
	\begin{align*}
		M_i\leq  \frac{\|\bar{B}_*^i\|\sigma_{\nu}\sqrt{n(n+d_i)\log\big(\frac{1+TY^*/\lambda(n+d_i)}{\delta}\big)}+\sigma_{\omega}\sqrt{n(n+d_i)\log\big(\frac{1+TY^*/\lambda(n+d_i)}{\delta}\big)}+\lambda^{1/2}s^i}{Y^*}:=M_{max}^i.
	\end{align*}
	Then we can upper bound $G_i$ as follows
	\begin{align*}
	G_i\leq \mathcal{C} \bigg(16\vee\frac{4S^2M_{max}^{i2}}{(n+d_i)U^i_0}\bigg):=\bar{G}_i
	\end{align*}
	Hence, the state norm is bounded as follows:
	\begin{align*}
	||x_t||\leq \mathcal{D}^i_1\sqrt{\beta^i_t(\delta)} \log (t) X^{\frac{n+d_i}{n+d_i+1}}+\mathcal{D}^i_2\sqrt{\log \frac{t}{\delta}}.
	\end{align*}

Adopted from \cite{abbasi2011regret}, it yields 
	\begin{align*}
	X_t\leq \big(\mathcal{D}^i_1\sqrt{\beta^i_t(\delta)} \log (t)+\mathcal{D}^i_2\sqrt{\beta^i_t(\delta)} \log (t) +\mathcal{D}^i_3\sqrt{\log \frac{t}{\delta}}\big)^{n+d_i}
	\end{align*}
	where
	
	\begin{align*}
	 &\mathcal{D}^i_1:=\frac{4}{1-\Upsilon_i}\big(\frac{\eta_i}{\Upsilon_i}\big)^{n+d_i}\bar{G}_i (1+2{\kappa^i}^2)^{\frac{n+d_i}{2(n+d_i+1)}}\\
	  \quad &  \mathcal{D}^i_2:=\frac{4}{1-\Upsilon_i}\big(\frac{\eta_i}{\Upsilon_i}\big)^{n+d_i}\bar{G}_i 2^{\frac{n+d_i}{2(n+d_i+1)}}\mathcal{V}_T^i\\
	  \quad & \mathcal{D}^i_3:=\frac{n\sqrt{2}}{1-\Upsilon_i}\big(\frac{\eta_i}{\Upsilon_i}\big)^{n+d_i}\sigma_{\omega}
	\end{align*}
	By elementary but tedious calculations one can show
\begin{align*}
\sqrt{\beta_t^i(\delta)}\leq 2n\sigma_{\omega}\log\frac{1}{\delta}+\sigma_{\omega}\sqrt{\lambda}s^i+n\sigma_{\omega}\log \frac{\det(V^i_t)}{\det(\lambda I_{(n+d_i)\times(n+d_i)})}
	\end{align*}

An upper bound for the third term in the right hand side is given by applying Lemma \ref{lemma:estimationerror_timeMinumum}. Letting $a_t=X_t^{1//(n+d_i+1)}$ and $c=\max\{1,\max_{1\leq s\leq t}||a_s||\}$ results in
	\begin{align*}
	& n\sigma_{\omega}\log \frac{\det(V^i_t)}{\det(\lambda I_{(n+d_i)\times(n+d_i)})}\leq n\sigma_{\omega}(n+d_i)\bigg(\log\frac{(n+d_i)\lambda+2{\mathcal{V}^i_t}^2}{(n+d_i)\lambda}t+\log\frac{(1+2{\kappa^i}^2){c}^2}{(n+d_i)\lambda}t\bigg)\\
	& \quad \leq n\sigma_{\omega}(n+d_i)\bigg(\log\frac{(n+d_i)\lambda+2{\mathcal{V}^i_t}^2}{(n+d_i)\lambda}t+\log\frac{(1+2{\kappa^i}^2)}{(n+d_i)\lambda}t\bigg)+2n\sigma_{\omega}(n+d_i)(n+d_i+1)\log^2 c
	\end{align*}
	By applying elementary calculations, it yields
	\begin{align*}
	c\leq \bar{L}^i+\bar{K}^i\log^2c
	\end{align*}
	where 
\begin{align*}
&\bar{L}^i=(\mathcal{D}^i_1+\mathcal{D}^i_2)\big(2n\sigma_{\omega}\log\frac{1}{\delta}+\sigma_{\omega}\sqrt{\lambda}s^i\big)\log t +\mathcal{D}^i_3
\sqrt{\log t/\delta}+\\
&\quad (\mathcal{D}^i_1+\mathcal{D}^i_2)n\sigma_{\omega}(n+d_i)\bigg(\log\frac{(n+d_i)\lambda+2{\mathcal{V}^i_t}^2}{(n+d_i)\lambda}t+\log\frac{(1+2{\kappa^i}^2)}{(n+d_i)\lambda}t\bigg)\log t\\
&\quad \bar{K}^i=2(\mathcal{D}_1^i+\mathcal{D}_2^i)n\sigma_{\omega}(n+d_i)(n+d_i+1)\log t.
\end{align*}
Using the property $\log x\leq x^s/s$ which holds $\forall s\in R^{+}$ and letting $s:=1$, one can write 
\begin{align*}
c\leq \frac{-\bar{L}_i+\sqrt{\bar{L}_i^2+4\bar{K}_i}}{2\bar{K}_i}.
\end{align*}
By elementary calculations first statement of the lemma is shown. The proofs of statements 2 and 3 are immediate and we skip them for the sake of brevity.
\end{pf}

\subsection{Proofs of Regret Bound Analysis Section}
\begin{pf} (Proof of Lemma \ref{lemma:R_2Bound})\label{prooflemma:R_2Bound}
Note that except the times instances that there is switch in policy, most terms in RHS of (\ref{eq:R2}) vanish. Denote covariance matrices of central ellipse and actuating mode $i^*$ by $V$ and $V^{i^*}$ respectively, and suppose at time steps $t_{n_1},...,t_{n_N}$ the algorithm changes the policy. Therefore, it yields $\det(V^{i^*}_{t_{n_i}}) \geq 2\det(V^{i^*}_{t_{n_{i-1}}})$ for $t\leq T^{i^*}_c$ and $\det(V_{t_{n_j}}) \geq 2\det(V_{t_{n_{j-1}}})$ for $t\geq T^{i^*}_c$. This results in
\begin{align}
\det(V^{i^*}_{T^{i^*}_c})&\geq 2^{N_1}\lambda ^{n+d_{i^*}}\label{eq:firstPhase}
\\
\det(V_T)&\geq 2^{N_2}\det(V_{T^{i^*}_c+1})\label{eq:secondPhase}
\end{align}
where $N_1$ and $N_2$ are number of switches in policy while actuating in the mode $i^*$ and fully actuating mode respectively. 
On one hand $\det(V^{i^*}_{T^{i^*}_c})\leq \lambda_{\max}^{n+d_{i^*}}(V^{i^*}_{T^{i^*}_c})$ where
\begin{align}
\nonumber \lambda_{\max}(V^{i^*}_{T^{i^*}_c})&\leq \lambda +\sum_{t=0}^{T^{i^*}_c-1}||z^{i^*}_t||^2\leq\\
&\quad\lambda +\big((1+2{\kappa^i}^2)\max_{0\leq t\leq T}\|x_t\|^2+2\max_{0\leq t\leq T}\|\nu_t(\mathcal{B}_i)\|^2\big)T^{i^*}_c\\
&\quad \lambda +{c^{i^*}}^2(1+2{\kappa^i}^2)(n+d_{i^*})^{2(n+d_{i^*})}T^{i^*}_c+4\sigma^2_{\nu}d_{i^*}\log\frac{8d_{i^*}T^{i^*}_c}{\delta}T^{i^*}_c
\end{align}
in which we applied the bound 
\begin{align}
\|\nu_t\|\leq \sigma_{\omega}\sqrt{2d_{i^*}\log\frac{d_{i^*}t}{\delta}}.
\end{align}

Using (\ref{eq:firstPhase}) $N_1$ is upper-bounded by
\begin{align}
N_1\leq \log_2 \bigg(1 +\frac{{c^{i^*}}^2(1+2{\kappa^i}^2)(n+d_{i^*})^{2(n+d_{i^*})}T^{i^*}_c+4\sigma^2_{\nu}d_{i^*}\log\frac{8d_{i^*}T^{i^*}_c}{\delta}T^{i^*}_c}{\lambda} \bigg)^{n+d_{i^*}}
\end{align}
On the other hand we have $\det(V_{T})\leq \lambda_{\max}^{n+d}(V_{T})$ where
\begin{align}
\nonumber\lambda_{\max}(V_{T})&\leq \lambda +\sum_{t=0}^{T-1}||z_t||^2\\
&\quad \leq\lambda +\sum_{t=0}^{T^{i^*}_c}\|z^{i^*}_t\|^2 +\sum_{t=T^{i^*}_c+1}^{T-1}\|z_t\|^2\\
&\quad \leq \lambda+\sum_{t=0}^{T_c^{i^*}}(1+2{\kappa^i}^2)\max_{0\leq t\leq T_c^{i^*}}\|x_t\|^2+2\max_{0\leq t\leq T_c^{i^*}}\|\nu_t(\mathcal{B}_i)\|^2+\sum_{t=T^{i^*}_c+1}^{T-1}\|z_t\|^2
\end{align}
Furthermore, 
\begin{align}
\Lambda:=\max _{0\leq t\leq T^{i^*}_c}\|\nu_t(\mathcal{B}_i)\|\leq \sigma_{\nu}\sqrt{2d_{i^*}\log(d_{i^*}T^{i^*}_c/ \delta) }\label{eq:Lambda} 
\end{align}
with probability at least $1-\delta/2$ together with upper-bounds of state norm in both initial exploration and optimism parts yields
\begin{align}
\nonumber\lambda_{\max}(V_T)\leq &\bar{\lambda}:=\lambda + {c^{i^*}}^2T^{i^*}_c(n+d_{i^*})^{2(n+d_{i^*})}(1+2{\kappa^i}^2)\\
\nonumber&\quad+4\sigma^2_{\nu}d_{i^*}\log(d_{i^*}T^{i^*}_c/ \delta)\\
&\quad+(T-T^{i^*}_c)\frac{32n\sigma^2_{\omega}(1+\kappa^2)}{(1-\Upsilon)^2} \log\frac{n(T-T^{i^*}_c)}{\delta}
\end{align}
Considering (\ref{eq:lowerBound}) we have 
\begin{align}
\lambda_{\min}(V_{T^{i^*}_c+1})\geq \sigma_{\star}^2 (T^{i^*}_c+1)\label{eq:minimumEigBound}
\end{align}
 now applying $\det(V_T)\leq \lambda_{\max}^{(n+d)}$ results in
\begin{align}
N_2\leq \log_2 \big(\frac{\bar{\lambda}}{\sigma_{\star}^2 (T_c^{i^*}+1)}\big)^{n+d}
\end{align}
Considering switch from IExp to SOFUA that can cause switch in the policy, the total number of switch in policy is $N_1+N_2+1$.
Now, applying  bounds of state norm for $t\leq T^{i^*}_c$ and $t> T^{i^*}_c$ complete proof.
\end{pf}
The following lemma adapts the proof of \cite{abbasi2011regret} to our setting which will be useful in bounding $R_3$.

\begin{lem}  (\cite{abbasi2011regret})\label{lemma:preliminary_for_R3Bounding}
On the event $F^{i^*}_{T_c^{i^*}}\cap E_{T_c^{i^*}}$ for $t\leq T_c^{i^*}$ and $F^{opt}_T\cap E_T$ for $t\geq T_c^{i^*}$, the following holds,
\begin{align}
\nonumber &\sum_{t=0}^{T}\|(\Theta_*-\tilde{\Theta}_t)^Tz_t\|^2\leq \frac{16(1+\kappa^2)}{\lambda}\\
\nonumber &\quad\times \bigg(2X_e^2 \beta^{i^*2}_{T_c^{i^*}}(\delta)\log\frac{\det(V^{i^*}_{T_c^{i^*}})}{\det{\lambda I}}\\
&\quad +X_c^2 \beta_T^2(\delta)\log \frac{\det(V_T)}{\det(V_{T_c^{i^*}})}\bigg)+2S^2\Lambda^2T_c^{i^*}
\end{align}
where $X_c^2=\frac{32n\sigma^2_{\omega}(1+\kappa^2)}{(1-\Upsilon)^2}\log \frac{n(T-T_c^{i^*})}{\delta}$and $X_e^2=c_c^{i^*}(n+d_{i^*})^{2(n+d_{i^*})}$
\end{lem}

\begin{pf}
Let $s^i_t=(\Theta^i_*-\tilde{\Theta}^i_t)^\top z^i_t$, for both $i=1$ and $i=i^*$, then one can write
\begin{align}
\|s^i_t\|\leq \|(\Theta^i_*-\hat{\Theta}^i_t)^\top z^i_t\|+\|(\hat{\Theta}^i_t-\tilde{\Theta}^i_t)^\top z^i_t\| \label{eq:decomp1}
\end{align}
For $t> T^{i^*}_c$ one can write
\begin{align}
 \|(\Theta-\hat{\Theta}_t)^\top z_t\|&\leq \|{V_t}^{1/2}(\Theta-\hat{\Theta}_t)\|z_t\|\|_{{V_t}^{-1}}\label{eq:CS}\\
 &\leq \|{V_{\tau}}^{1/2}(\Theta-\hat{\Theta}_t)\| \|z_t\|_{{V_t}^{-1}}\sqrt{\frac{\det(V_t)}{\det(V_{\tau})}}\label{eq:Lemma 11Abbasi}\\
 &\leq \sqrt{2}\|{V_{\tau}}^{1/2}(\Theta-\hat{\Theta}_t)\|\|z_t\|_{{V_t}^{-1}}\label{eq:policyUpdate}\\
  &\leq \sqrt{2\beta_{\tau}(\delta)} \|z_t\|_{{V_t}^{-1}}\label{eq:lamMax}
\end{align}
where $\tau\leq t$ is the last time that policy change happened. We applied Cauchy-Schwartz inequality in (\ref{eq:CS}). The inequality (\ref{eq:Lemma 11Abbasi}) follows from the Lemma 11 in \cite{abbasi2011regret}. Furthermore, applying the update rule gives (\ref{eq:policyUpdate}) and finally (\ref{eq:lamMax}) is obtained using the property $\lambda_{\max}\leq Tr(M)$ for $M\succeq 0$. 

Now, for $t\geq T^{i^*}_c$ we have 

\begin{align*}
 \|s_t\|^2\leq 8 \beta^i_{\tau}(\delta) \|z^i_t\|_{{V_t^i}^{-1}}^2  
\end{align*}

which in turn is written

\begin{align}
\nonumber \sum_{t=T_c}^{T}\|(\Theta_*-\tilde{\Theta}_t)^\top z_t\|^2&\leq \frac{8X^2_c(1+\kappa^2)\beta_T(\delta)}{\lambda}\\
\nonumber&\times\sum_{t=T^{i^*}_c+1}^{T}\min\{\|z_t\|^2_{V_t^{-1}}, 1\}\\
 & \leq\frac{16X^2_c(1+\kappa^2)\beta_T(\delta)}{\lambda}\log\frac{\det(V_T)}{\det(V_{T^{i^*}_c})} \label{eq:t>T_cSum}
\end{align}

where in the second inequality we applied the Lemma 10 of \cite{abbasi2011regret}.

For $t\leq T^{i^*}_c$ the algorithm applies an extra exploratory noise, we still have same decomposition (\ref{eq:decomp1}) with $i=i^*$. However, to upper-bound $\|s_t^{i^*}\|$ we need to some algebraic manipulation which is substituting $\bar{z}^{i^*}_t$ in terms of $z^{i^*}_t$. Bringing this into mind, by following similar steps as of (\ref{eq:CS}-\ref{eq:lamMax}) it yields 

\begin{align}
 \nonumber\|(\Theta^{i^*}-\hat{\Theta}^{i^*}_t)^\top \bar{z}^{i^*}_t\|&\leq \|(\Theta^{i^*}-\hat{\Theta}^{i^*}_t)^\top (\bar{z}^{i^*}_t+\xi_t-\xi_t)\|\\
  \nonumber &\leq\|(\Theta^{i^*}-\hat{\Theta}^{i^*}_t)^\top z^{i^*}_t\|+\|(\Theta^{i^*}-\hat{\Theta}^{i^*}_t)^\top \xi_t\|\\
  \nonumber &\leq\|V_t^{i^*1/2}(\Theta^{i^*}-\hat{\Theta}^{i^*}_t)\|z^{i^*}_t\|\|_{V_t^{i^*-1}}+S\Lambda\\
 \nonumber&\leq \|V_{\tau}^{i^*1/2}(\Theta^{i^*}-\hat{\Theta}^{i^*}_t)\| \|z^{i^*}_t\|_{V_t^{i^*-1}}\\
 \nonumber &\times\sqrt{\frac{\det(V^{i^*}_t)}{\det(V^{i^*}_{\tau})}}+S\Lambda \\
 \nonumber&\leq \sqrt{2}\|V_{\tau}^{i^*1/2}(\Theta^{i^*}-\hat{\Theta}^{i^*}_t)\| \|z^{i^*}_t\|_{V_t^{i^*-1}}+S\Lambda\\
 \nonumber &\leq \sqrt{2\beta^{i^*}_{\tau}(\delta)} \|z^{i^*}_t\|_{V_t^{i^*-1}}+S\Lambda \end{align}

where $\xi=\begin{pmatrix} 0 \\ \nu_t(\mathcal{B}_i) \end{pmatrix}$. 
Applying this result gives

\begin{align}
\nonumber \sum_{t=0}^{T^{i^*}_c}\|(\Theta^{i^*}_*-\tilde{\Theta}^{i^*}_t)^\top \bar{z}_t^{i^*}\|^2&\leq \frac{16\big((1+2{\kappa^{i^*}}^2)X^2_e+2\Lambda\big){\beta^{i^*}_{T^{i^*}_c}}(\delta)}{\lambda}\\
\nonumber&\times\sum_{t=0}^{T^{i^*}_c}\min\{\|z^{i^*}_t\|^2_{V_t^{{i^*}-1}}, 1\}+8S^2\Lambda^2T^{i^*}_c\\
 \nonumber& \leq\frac{32\big((1+2{\kappa^{i^*}}^2)X^2_e+2\Lambda\big)\beta^{i^*}_{T^{i^*}_c}(\delta)}{\lambda}\log\frac{\det(V^{i^*}_{T^{i^*}_c})}{\det(\lambda I)}\\
 &+8S^2\Lambda^2T^{i^*}_c\label{eq:t<T_cSum}
 \end{align}

where in first inequality we applied (\ref{eq:boundzn}) and (\ref{eq:Lambda}).   Similar to (\ref{eq:t>T_cSum}) in the second inequality we applied the Lemma 10 of \cite{abbasi2011regret}.
 Combining (\ref{eq:t>T_cSum}) and (\ref{eq:t<T_cSum}) completes proof.
\end{pf}

\begin{pf}(Proof of Lemma \ref{lemma:R_3Bound})\label{prooflemma:R_3Bound})
In upper-bounding the term $R_3$ we skip a few straight forward steps which can be found in \cite{abbasi2011regret} and \cite{lale2020explore} and write

\begin{align}
\nonumber&|R_3|\leq \bigg(\sum_{t=0}^{T^{i^*}_c}\bigg\|P(\tilde{\Theta}^{i^*}_t, Q_*, R_*^{i^*})^{1/2}\big(\tilde{\Theta}^{i^*}_t-\Theta^{i^*}_*\big)^\top \bar{z}^{i^*}_t\bigg\|^2\bigg)^{1/2}\\
\nonumber&\times \bigg(\sum_{t=0}^{T^{i^*}_c}\big(\big\|P(\tilde{\Theta}^{i^*}_t, Q_*, R_*^{i^*})^{1/2}
(\tilde{\Theta}_t^{i^*})^\top\bar{z}^{i^*}_t\big\|+\big\|P(\tilde{\Theta}^{i^*}_t, Q_*, R_*^{i^*})^{1/2}{\Theta_*^{i^*}}^\top \bar{z}^{i^*}_t\big\|\big)^2\bigg)^{1/2}\\
\nonumber&+\bigg(\sum_{t=T^{i^*}_c}^{T}\bigg\|P(\tilde{\Theta}_t, Q_*, R_*)^{1/2}\big(\tilde{\Theta}_t-\Theta_*\big)^\top z_t\bigg\|^2\bigg)^{1/2}\\
\nonumber&\times \bigg(\sum_{t=T^{i^*}_c}^{T}\big(\big\|P(\tilde{\Theta}_t, Q_*, R_*)^{1/2}\tilde{\Theta}_t^\top z_t\big\|+\big\|P(\tilde{\Theta}_t, Q_*, R_*)^{1/2}\Theta_*^\top z_t\big\|\big)^2\bigg)^{1/2}\\
 \nonumber&\leq \sqrt{\frac{32D\big((1+2{\kappa^{i^*}}^2)X^2_e+2\Lambda\big)\beta^{i^*}_{T^{i^*}_c}(\delta)}{\lambda}\log\frac{\det(V^{i^*}_{T^{i^*}_c})}{\det(\lambda I)}+8DS^2\Lambda^2T^{i^*}_c}\\
 & \times \sqrt{4T^{i^*}_cDS^2\big((1+2{\kappa^{i^*}}^2)X^2_e+2\Lambda\big)}\label{eq:first}\\
 \nonumber&+\sqrt{\frac{16D(1+\kappa^2)\beta^2_T(\delta)X_c^2}{\lambda}\log\frac{\det(V_T)}{\det(V_{T^{i^*}_c})}}\\
 &\times\sqrt{4(T-T^{i^*}_c)DS^2(1+\kappa^2)X_c^2}\label{eq:second}
 \end{align}

where in the inequalities (\ref{eq:first}) and (\ref{eq:second}) we applied (\ref{eq:t<T_cSum}) and (\ref{eq:t>T_cSum}) (from the Lemma \ref{lemma:preliminary_for_R3Bounding} ) respectively.
The remaining step is following upper-bounds

\begin{align*}
&\log\frac{\det(V^{i^*}_{T^{i^*}_c})}{\det(\lambda I)}\leq (n+d_{i^*})\log\big(1+\frac{\big((1+2{\kappa^{i^*}}^2)X^2_e+2\Lambda\big)}{\lambda (n+d_{i^*})}\big)\\
&\log\frac{\det(V_{T})}{\det(V_{T^{i^*}_c})}\leq (n+d)\\
&\times\log \frac{\lambda(n+d)+T^{i^*}_c\big(2(1+2{\kappa^{i^*}}^2)X^2_e+4\Lambda+2\bar{\Lambda}\big)+(T-T^{i^*}_c)(1+\kappa^2)X_c^2}{\sigma_{\star}^2 T^{i^*}_c(n+d)}
\end{align*}
\normalsize
where 
\begin{align}
\bar{\Lambda}:= \sigma_{\nu}\sqrt{2(d-d_{i^*})\log((d-d_{i^*})T^{i^*}_c/ \delta) }
\end{align}

and in the second inequality we applied  (\ref{eq:minimumEigBound}). Considering the definitions of $X_e^2$, $X_c^2$ and $\Lambda$ we can easily notice the statement of lemma holds true.
\end{pf}

\end{document}